\documentclass[prb,twocolumn,showpacs]{revtex4}
\usepackage{graphicx}
\usepackage{amssymb}
\usepackage{amsmath}

\newcommand{\ii}{\mathrm{i}}

\let \Im \relax
\DeclareMathOperator{\Im}{Im}

\begin{document}

\title{Chebyshev approach to quantum systems coupled to a bath}

\author{Andreas Alvermann}
\author{Holger Fehske}
\affiliation{
Institut f{\"u}r Physik, Ernst-Moritz-Arndt-Universit{\"a}t Greifswald,
17489 Greifswald, Germany }

\begin{abstract}
We propose a new concept for the dynamics of a quantum bath, the
Chebyshev space, and a new method based on this concept,
the Chebyshev space method.
The Chebyshev space is an abstract vector space that exactly represents
the fermionic or bosonic bath degrees of freedom,
without a discretization of the bath density of states.
Relying on Chebyshev expansions the Chebyshev space
representation of a bath has very favorable properties with respect
to extremely precise and efficient calculations of groundstate
properties, static and dynamical correlations, and time-evolution
for a great variety of quantum systems.
The aim of the present work is to introduce the Chebyshev space
in detail and to demonstrate the capabilities
of the Chebyshev space method.
Although the central idea is derived in full generality the focus is
on model systems coupled to fermionic baths.
In particular we address quantum impurity problems, such as
an impurity in a host or a bosonic impurity with a static barrier,
and the motion of a wave packet on a chain coupled to leads.
For the bosonic impurity, the phase transition from a
delocalized electron to a localized polaron in arbitrary dimension is
detected. For the wave packet on a chain, we show how the Chebyshev
space method implements different boundary conditions,
including transparent boundary conditions replacing infinite
leads.
Furthermore the self-consistent solution of the Holstein model in infinite
dimension is calculated.
With the examples we demonstrate how highly accurate results for system
energies, correlation and spectral functions, and time-dependence of
observables are obtained with modest computational effort.
\end{abstract}


\maketitle

\section{Introduction}

The calculation of spectral or dynamical properties of quantum
systems, expressed through spectral functions or captured in the
time evolution of a wave function,
is one of the most important and most promising applications
of modern numerical techniques in theoretical physics or chemistry.
For many purposes,
approximation free techniques that allow to calculate numerically exact
results for arbitrary Hamiltonians on large Hilbert spaces are of interest.
One of the most powerful tools in this context are techniques based on
Chebyshev expansions, like the kernel polynomial method (KPM)~\cite{SR94,WZ94,Wa94,SRVK96,WWAF06},
which yield results of high accuracy with modest computational effort.
Chebyshev techniques often outperform the Lanczos algorithm
with respect to accuracy and efficiency, and are applicable to problems
that are beyond the reach of matrix diagonalization.
In this article we propose an extension of Chebyshev techniques
that considerably enlarges their field of applications.
Possible new applications we have in mind include 
(i) damping and decoherence in quantum systems coupled to an
environment like a bosonic heat bath,
(ii) transport through quantum systems coupled to fermionic baths,
(iii) the solution of quantum impurity models,
(iv) non-equilibrium dynamics of mesoscopic devices coupled to leads,
and in general,
(v) the calculation of static and dynamical correlation functions or
time propagation in these physical situations,
(vi) the combination with diagrammatic Green function techniques,
(vii) the treatment of degrees of freedom with non-trivial
dynamics like phonons with dispersion.

We concentrate here on a situation for which the development of ideas
is particularly clear: A quantum system
coupled to a fermionic bath. 
The bath serves as a reservoir for fermions, which can hop from the
bath to the quantum system and back.
The standard example is that of a mesoscopic system
contacted with leads.~\cite{Da95_2,FG97}
For a single quantum dot, the appropriate model is the famous Anderson
model for the Kondo effect~\cite{He97},
which describes an impurity site with Coulomb interaction embedded in
a host of non-interacting electrons.
This model also arises in dynamical mean-field theory (DMFT) where the
solution of an Anderson model in dependence on a host spectral
function is a central issue.~\cite{GKKR96}

In all these cases the quantum system allows for many-particle interactions, which
give rise to non-trivial correlations.
For their description one has to rely on several
correlation functions whose calculation requires the full apparatus
of many-body theory.
In the numerical calculation we have to deal with complicated objects
like a Fock space of many fermions.
Due to the coupling between the quantum system and the bath,
further correlations between the quantum system and the bath develop.
As a consequence one cannot separate the degrees of freedom of the
quantum system and of the bath, but must treat their evolution in
parallel.
However, the bath consists of non-interacting fermions,
and is entirely described by single-particle spectral functions,
as no  correlations exist in absence of coupling to the
quantum system.
Plainly spoken, we do not need to describe all details of the bath,
but only how its presence influences the dynamics of the quantum system.
It is important to exploit this simplification to make a calculation possible.
Trivial as this point may seem, it is not easily incorporated into a
numerical calculation.
The numerically exact techniques to which Chebyshev techniques
belong require an explicit representation of the bath degrees of
freedom in terms of a Hamiltonian and an associated Hilbert space.
In the worst case such a representation again involves a many-body Fock
space, describing the bath with the same complexity as the interacting
quantum system.

The new idea we promote here 
provides a different solution which is based on Chebyshev expansions of
spectral functions of the bath.
The Chebyshev expansion supplies the information
necessary for an exact calculation of correlations function for the
quantum system coupled to the bath,
but avoids the introduction of redundant parameters that unnecessarily
complicate the calculation.
Thereby, the simplification of a non-interacting bath is exploited
while the full coupled dynamics of the quantum system and the bath is
retained.
The realization of this idea amounts to the construction of an
abstract vector space, which we call the Chebyshev space (CS).
The combination of the CS with computational techniques based on
Chebyshev expansions, in particular KPM, will be referred to as the Chebyshev space method~(CSM).

\begin{figure}
\includegraphics[width=0.5\textwidth]{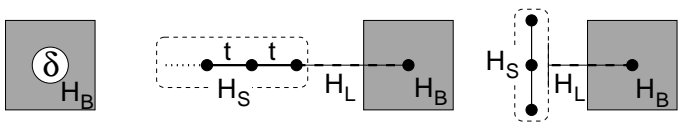}
\caption{Different system-bath geometries used for examples in this
  article. On the left, a single site is embedded in a host (cf.
  Eq.~\eqref{HOneSite}), e.g. an impurity in a lattice.
In the middle, a chain is coupled to a bath
(cf. Eq.~\eqref{HamChain}), allowing an electron to move along the chain
and to hop to the bath and return.
On the right, a single site with local bosonic degrees of freedom is coupled
to a bath (cf. Eq.~\eqref{HamHol}).
An electron can excite bosons at the site, and hop to the bath while
the bosons are left behind.
The last two cases fit into the scheme of Eq.~\eqref{HamGeneral},
where the coupling between system and bath is given by a
Hamiltonian $H_L$.
}
\label{Fig:Bath}
\end{figure}

We demonstrate the CSM in this article for a number of examples
(cf. Fig.~\ref{Fig:Bath}),
including (i) the calculation of spectral functions and groundstate energy
for an impurity in a host in various dimensions,
(ii) the description of the electron-polaron phase transition for a
bosonic impurity, 
(iii) the self-consistent solution of the Holstein model within DMFT,
and (iv) the time propagation of a wave packet on a chain with different
boundaries conditions.

We do not include examples for interacting fermions at finite density
that involve renormalization of the bath due to the creation of
particle-hole pairs like in the Kondo problem.
The discussion would exceed a tolerable length of this article
and obscure the presentation of the new ideas.
We return to this issue in the conclusion,
and concentrate here on examples better suited for the introduction of
the technical details of the CS construction.

The article is organized as follows.
In Sec.~\ref{Sec:Cheby}, we remind the reader 
of Chebyshev expansions for spectral functions,
and fix some notations used in the remainder.
In Sec.~\ref{Sec:General} we introduce the CS,
and explain in Sec.~\ref{Sec:Host} the implementation of the CSM 
for the example of an impurity embedded in a host.
In Sec.~\ref{Sec:TruncatedHB} we derive an important property of the
CS and discuss its practical relevance.
Sec.~\ref{Sec:Chain} describes how self-consistent calculations are
performed in CSM, and Sec.~\ref{Sec:Concurrent} 
demonstrates how the concurrent dynamics of a quantum system and a bath,
when degrees of freedom evolve in parallel, is accounted for.
This provides the basis for the study of the phase transition for
a bosonic impurity in Sec.~\ref{Sec:Bosonic},
and to the self-consistent solution of the Holstein model within DMFT in
Sec.~\ref{Sec:Holstein}.
In Sec.~\ref{Sec:Time} we address the time evolution of a wave packet
on a long chain with different boundary conditions realized by CSM.
We summarize in Sec.~\ref{Sec:Summary} and point out possible future
advancements of CSM. 
In the Appendices, we provide a short account of technical aspects of
Chebyshev expansions and KPM, and the
derivation of some mathematical results used in the text.

\section{Chebyshev expansions and the kernel polynomial}
\label{Sec:Cheby}

The recurrent theme of this article is the expansion of a spectral
function in a series of Chebyshev polynomials.
To make the article self-contained we add an appendix on 
technical issues of Chebyshev expansions and KPM (App.~\ref{App:Cheby}).
For a more detailed exposition, we refer the reader to the
recent review Ref.~\onlinecite{WWAF06}, and references cited therein.

The Chebyshev polynomials $T_n(x)$ are defined by the two-term
recurrence~\cite{AS70}
\begin{equation}\label{ChebyRecursion}
\begin{split}
  T_0(x) = 1 \;, \quad T_1(x)=x \;, \\
  T_{n+1}(x) = 2xT_n(x) - T_{n-1}(x) \;,
\end{split} 
\end{equation}
which is equivalent to $T_n(x) = \cos (n \arccos x)$.
The Chebyshev polynomials are mutually orthogonal on the interval
$[-1,1]$ with respect to the scalar product given by the weighting function
$(1-x^2)^{-1/2}$,
obeying the relations
\begin{equation} \label{ChebyOrthogonal}
  \int_{-1}^1 \frac{T_m(x) T_n(x)}{\pi \sqrt{1-x^2}} dx = 
  \begin{cases}
    1 \;, &n=m=0 \\
    \frac{1}{2} \delta_{mn} \;, &n,m \ne 0
  \end{cases} \;.
\end{equation}

Eq.~\eqref{ChebyOrthogonal} implies that the Chebyshev polynomials
form an orthogonal basis for functions defined on the interval
$[-1,1]$.
To expand a spectral function, defined as
\begin{equation}\label{ADef}
  A(\omega) = \langle \psi|
  \delta(\omega-H) |\psi \rangle =
  -\frac{1}{\pi} \lim_{\eta \to 0^+} \Im \langle \psi|
  [\omega+\ii\eta-H]^{-1} |\psi \rangle
\end{equation}
 to some Hamiltonian $H$ and vector $|\psi\rangle$,
in a series of Chebyshev polynomials 
it must therefore vanish outside $[-1,1]$.
This can be achieved by re-scaling the Hamiltonian $H$ as
$H = p \tilde{H} +q$,
where $p,q$ is chosen in such a way that all eigenvalues of
$\tilde{H}$ lie in $[-1,1]$, or
equivalently $\|\tilde{H}\|<1$.
After a corresponding variable substitution $x=(\omega-q)/p$,
the Chebyshev expansion for $A(\omega)$ reads
\begin{equation}\label{ACheby}
  A(\omega) =\frac{1}{p} \, \frac{1}{\pi \sqrt{1-x^2}}   \Big[ \mu_0 + 2 \sum_{n=1}^\infty \mu_n T_n(x)
  \Big] \;,
\end{equation}
for $A(\omega)$ on the interval
$[\omega_\mathrm{min},\omega_\mathrm{max}]$
with $\omega_\mathrm{min}=-p+q$, $\omega_\mathrm{max}=p+q$.
Comparing this expansion to the definition~\eqref{ADef}, and using the
orthogonality relation~\eqref{ChebyOrthogonal}, we find that the
coefficients in this series -- the Chebyshev moments -- are given by 
\begin{equation}\label{ChebyMoms}
  \mu_n = \int_{-1}^1 T_n(x) A(px+q) dx = \langle \psi|
  T_n(\tilde{H}) |\psi \rangle \;.
\end{equation}
We will assume in the following that any
Hamiltonian is properly scaled to
$[-1,1]$.
The introduction of scaling factors
is straightforward most times.
Whenever important, we will explicitly discuss the
consequences of scaling.

The calculation of the vectors $T_n(\tilde{H}) |\psi \rangle$ 
in Eq.~\eqref{ChebyMoms}
can be accomplished by means of the two-term
recurrence~\eqref{ChebyRecursion}.
Starting with the vector $|\psi\rangle$,
one iteratively obtains the $(n+1)^{th}$ vector from the $n^{th}$ and
$(n-1)^{th}$ vector. 
The calculation of the moments therefore requires
(i) to apply $H$ to a vector, i.e. to perform matrix-vector
multiplication if $H$ is explicitly given as a matrix, and (ii) to
evaluate a scalar product of two vectors.

In any practical application only a finite number $N$ of moments can
be calculated,  and $A(\omega)$ has to be reconstructed from a truncated
series, taking the first $N$ terms
in Eq.~\eqref{ACheby}.
The reconstruction of $A(\omega)$ from such a finite series 
is the issue of KPM (see App.~\ref{App:Cheby}).
High resolution of $A(\omega)$ 
is already obtained with a fairly small number of moments,
thus allowing for accurate calculations with moderate demands on
computational time or memory.
The resolution of KPM scales linearly with $N$.
It is usually much
better than for the Lanczos (Recursion) Method~\cite{HHK72,GB87,ZSP94}, 
and can be unrestrictedly increased without problems~\cite{WWAF06}.

\section{The general scheme}\label{Sec:General}

A coupled quantum
system and bath is generally described by a
Hamiltonian of the form~\cite{Da95_2,Mah00}
\begin{equation}\label{HamGeneral}
  H = H_S + H_L + H_B \;.
\end{equation}
$H_S$ and $H_B$ denote the
Hamiltonian for the quantum system and the bath, respectively,
and $H_L$ describes the coupling
between system and bath.
Without coupling, for $H_L=0$, the degrees of
freedom of the quantum system and
the bath evolve independently, i.e.
$H_S$ and $H_B$ commute.
We restrict ourselves in this article to situations with a single
spinless fermion. 
This allows to introduce the CS
without a discussion of complications inherent to many-fermion physics.
The CSM can be extended to finite fermion
density, or the inclusion of spin degrees of freedom,
with the same implementation of a CS as given in this article.
We return to that issue in Sec.~\ref{Sec:Summary}.

Since the bath consists
of non-interacting fermions, $H_B$
is a bilinear Hamiltonian
\begin{equation}\label{HamBath}
  H_B = \sum_{\alpha\beta} (H_B)_{\alpha\beta} f^\dagger_\alpha
  f_\beta \;,
\end{equation}
where $f^{(\dagger)}_\alpha$ are
fermionic operators for the bath
degrees of freedom.
The indices
$\alpha,\beta$ can denote e.g. an arbitrary set of
orbitals, sites of a lattice, or, if $(H_B)_{\alpha\beta}$ is
diagonal, the eigenstates of $H_B$. 
A change of indices corresponds to a
unitary transformation of the matrix $(H_B)_{\alpha\beta}$.
The system Hamiltonian $H_S$ can be of arbitrary form,
involving any type of interaction. 
We do not specify $H_S$ now, and discuss examples for
various different $H_S$ later.

More can be said about the coupling term $H_L$.
In general, the bath is in contact to one (or a few) site(s) of the
quantum system, say site $0$ with fermionic operator $c^{(\dagger)}_0$.
Fermions hop from this site to the bath, and back.
The appropriate choice for $H_L$ is 
\begin{equation}\label{HamLink}
  H_L =  \sum_\alpha t_\alpha (c^\dagger_0 f_\alpha + f^\dagger_\alpha
  c^{}_0)  \qquad (t_\alpha \in \mathbb{R}) \;,
\end{equation}
or a sum of terms of this form.
It is convenient to introduce a fermionic operator $d^{(\dagger)}$ by
\begin{equation}\label{DOp}
  d^\dagger = \sum_\alpha a_\alpha f^\dagger_\alpha  \;, \quad 
  a_\alpha = t_\alpha/(\sum_\alpha t_\alpha^2)^{1/2} \;.
\end{equation}
As $\sum_\alpha a_\alpha^2 = 1$,
the fermionic anticommutator relation $\{d,d^\dagger\}=1$ is fulfilled.
With the introduction of $d^{(\dagger)}$,
$H_L$ acquires the form of a hopping term between two sites,
\begin{equation}\label{HamLink2}
  H_L = V (c^\dagger_0 d + d^\dagger c_0 ) \;, \quad
  V = (\sum_\alpha t_\alpha^2)^{1/2} \;.
\end{equation}
Depending on the meaning of indices in Eq.~\eqref{HamBath},
$d$ can denote a concrete orbital or a site in a lattice,
or just an abstract linear combination of $f$-operators.
If we think of a mesoscopic system contacted to a lead,
then $c^{(\dagger)}_0$ and $d^{(\dagger)}$ denote operators for the
contact point in the system and lead, respectively.

We define the bath spectral function $A_B(\omega)$
as the spectral function of the $d$-orbital,
\begin{equation}
  A_B(\omega) = \langle \mathrm{vac}| d \, \delta(\omega-H_B) \, d^\dagger
  |\mathrm{vac} \rangle
\end{equation}
where $|\mathrm{vac}\rangle$ is the bath vacuum,
i.e. $d|\mathrm{vac}\rangle=0$.
Remember that we study situations with a single fermion. For finite
fermion density, we had to consider both the particle and hole part of
the spectral function, and to account for Pauli blocking.

After the transformation~\eqref{DOp},
the bath occurs in the calculation of correlation functions for the
quantum system only through the spectral function $A_B(\omega)$ and
the coupling strength $V$.
The precise coefficients in Eqs.~\eqref{HamBath},~\eqref{HamLink} do
not occur.
In this sense, all Hamiltonians~\eqref{HamGeneral} with the same
$A_B(\omega)$, V, but potentially different $H_B$, $H_L$, are equivalent.
Often the problem under study is 
defined in this way, by specifying $A_B(\omega)$ and $V$, without
an explicit representation of $H_B$ or $H_L$ as in
Eqs.~\eqref{HamBath},~\eqref{HamLink} at hand.

Given such a problem, how can it be accessed
within numerically exact techniques like Lanczos or
KPM?
Since these techniques need the Hamiltonian $H$ explicitly given,
it seems inevitable to use an explicit representation of $H_B$
in the form~\eqref{HamBath}.
One possible way to obtain such a representation is to discretize
$A_B(\omega)$ by a finite number of $\delta$-peaks, as
\begin{equation}\label{Aapproximation}
  A_B(\omega) \approx \sum_\alpha w_\alpha
  \delta(\omega-\epsilon_\alpha) \;.
\end{equation}
This approximation translates to $H_B = \sum_\alpha
\epsilon_\alpha f^\dagger_\alpha f_\alpha$, and 
$d^\dagger = \sum_\alpha w_\alpha^{1/2} f^\dagger_\alpha$.
A calculation then relies on our ability to construct a good
approximation~\eqref{Aapproximation}.
This poses certain questions:
How to choose the positions $\epsilon_\alpha$ of peaks, how to choose their
weights $w_\alpha$, and how many of them are needed to approximate
$A_B(\omega)$ sufficiently well?
When is an approximation to $A_B(\omega)$ sufficiently good,
and what is the precise meaning of ``$\approx$'' in Eq.~\eqref{Aapproximation}?
Is there an optimal way to choose $\epsilon_\alpha$, $w_\alpha$ for a given
number of peaks?
While the result of a calculation for the quantum system does not depend
on the precise form of $H_{B}$, it does depend on the approximation
for $H_B$ or $A_B(\omega)$. How can we control this dependence?
There is no definite answer to these questions,
and the need to discretize $A_B(\omega)$ is quite unsatisfactory.
Our new proposition, the use of the CS in the CSM, avoids the discretization of
$A_B(\omega)$.
It works without a representation of
$H_B$ in the form~\eqref{HamBath},
but addresses $H_B$ only via the spectral function $A_B(\omega)$.

The central ingredient of CS(M) is a representation of $H_B$ related to the
Chebyshev expansion of $A_B(\omega)$.
To obtain this representation, define the Chebyshev vectors
\begin{equation}\label{ChebyVectors}
  |n\rangle = T_n(H_B) d^\dagger |\mathrm{vac}\rangle
\end{equation}
for $n\ge 0$.
These vectors are neither normalized nor orthogonal.
We need only the scalar products
\begin{equation}\label{ChebyImpurity1}
  \langle 0|n\rangle = \langle\mathrm{vac}|d \, T_n(H_B) \,
  d^\dagger|\mathrm{vac}\rangle = \mu^B_n \;,
\end{equation}
where $\mu^B_n$ is the $n^{th}$ Chebyshev moment of the bath spectral
function $A_B(\omega)$, according to the expansion~\eqref{ACheby}.
The Chebyshev vectors span a Hilbert space $\mathcal{H}_c$,
the CS.
By definition, $\mathcal{H}_c$ is a subspace of the Fock space for the
bath operators $f^{(\dagger)}$,
but we refer to the Chebyshev vectors only as abstract vectors, whose
possible representation in terms of the $f^{(\dagger)}$ is irrelevant.

From the recurrence relation~\eqref{ChebyRecursion} 
it follows that the operation of $H_B$ on $\mathcal{H}_c$ is given by
\begin{equation}\label{ChebyImpurity2}
  H_B |n\rangle = \begin{cases}
    |1\rangle  \quad &,\; n = 0 \\
    (1/2)(|n-1\rangle + |n+1\rangle) &,\; n \ne 0
  \end{cases} \;.
\end{equation}
Since $|n\rangle$ is a one-fermion state, $d|n\rangle$ is proportional
to $|\mathrm{vac}\rangle$.
Together with the definition of $\mu^B_n$ this yields
\begin{equation}\label{ChebyImpurity3}
\begin{split}
d |n\rangle &= d\, T_n(H_B) d^\dagger |\mathrm{vac}\rangle \\
&= |\mathrm{vac}\rangle\langle\mathrm{vac}|  d\, T_n(H_B)d^\dagger
|\mathrm{vac}\rangle = \mu^B_n |\mathrm{vac}\rangle
\end{split}
\end{equation}
and 
\begin{equation}\label{ChebyImpurity4}
\begin{split}
  d^\dagger d |n\rangle &= d^\dagger d\, T_n(H_B) d^\dagger |\mathrm{vac}\rangle \\
&= d^\dagger|\mathrm{vac}\rangle\langle\mathrm{vac}|  d\, T_n(H_B)d^\dagger |\mathrm{vac}\rangle =
\mu^B_n |0\rangle \;.
\end{split} 
\end{equation}
The reader should note that in these equations the only parameters
are the Chebyshev moments $\mu^B_n$ of the spectral function
$A_B(\omega)$.

Eqs.~\eqref{ChebyImpurity1}--\eqref{ChebyImpurity4}
constitute the basis of our CS approach.
What we have achieved is that $H_B$ is put into a form that makes no
reference to a representation like~\eqref{HamBath}.
Still, $H_B$ is given as a Hamiltonian acting on a Hilbert space.
This is a very useful form for $H_B$,
which can be used within numerically exact techniques like Lanczos 
or KPM \textit{and} avoids discretization of $A_B(\omega)$.

\section{A fermionic site embedded in a host}\label{Sec:Host}

We illustrate the use of the CS(M)
with the example
\begin{equation}\label{HOneSite}
  H = - \Delta d^\dagger d + H_B
\end{equation}
of an unperturbed system $H_B$ with a perturbation
$-\Delta d^\dagger d$.
This is the Hamiltonian of a single site in a sea of non-interacting fermions,
like an impurity embedded in a host or lattice.
Note that this example is not of the form~\eqref{HamGeneral},
as it deals with a site embedded in the bath,
while in Eq.~\eqref{HamGeneral} and later examples
fermions leave the bath by hopping to the quantum system.

Our goal is to calculate the spectral function 
\begin{equation}\label{AImpurity}
  A(\omega) = \langle \mathrm{vac}| d \, \delta(\omega-H) \, d^\dagger
  |\mathrm{vac} \rangle
\end{equation}
to given $A_B(\omega)$ and
$\Delta$ within CSM, i.e. using KPM with the
Eqs.~\eqref{ChebyImpurity1}--\eqref{ChebyImpurity4} for the CS.
We can compare the results to the exact result
\begin{equation}\label{Gexact}
  G(z) = [G_B(z)^{-1} + \Delta ]^{-1}
\end{equation}
for the corresponding Green functions
\begin{equation}
  G_{(B)}(z) = \langle \mathrm{vac}| d [z
  -H_{(B)}]^{-1} d^\dagger |\mathrm{vac} \rangle \;.
\end{equation}

To obtain the Chebyshev moments $\mu_n$ of
$A(\omega)$, we have to recursively calculate the vectors
$T_n(H)d^\dagger|\mathrm{vac}\rangle$
according to Eq.~\eqref{ChebyRecursion}.
The calculation proceeds in the space $\mathcal{H}_c$,
and each vector is given as a linear combination
of Chebyshev vectors $|n\rangle$.
With $H_B$ according to~\eqref{ChebyImpurity2} and
$\Delta d^\dagger d$ according to~\eqref{ChebyImpurity4}, 
$H$ is represented by the matrix 
\begin{equation}\label{HamMatrix}
 (H)_{mn} = \frac{1}{2} \begin{pmatrix} -2 \Delta \mu^B_0 & 1 - 2\Delta \mu^B_1 & -
   2 \Delta \mu^B_2 & - 2 \Delta \mu^B_3 & \hdots  \\ 2 & 0 & 1 & 0 & \hdots \\ & 1 & 0
    & 1 \\ & & 1 & 0 \\ & & \vdots & & \ddots
  \end{pmatrix}
\end{equation}
with $H|n\rangle = \sum_m (H)_{mn} |m\rangle$.
Note that $(H)_{mn} \ne \langle m|H|n\rangle$,
because the Chebyshev vectors are not orthogonal.
Especially, $(H)_{mn}$ is not symmetric. Nevertheless, $H$ is
hermitian by definition (an explicit calculation is given in App.~\ref{App:Herm}). 

Starting with the vector $|0\rangle = d^\dagger
|\mathrm{vac}\rangle$,
in any step of the Chebyshev iteration we apply $H$ according to~\eqref{HamMatrix} to obtain 
the vector $T_n(H)|0\rangle$ from previous vectors
$T_{n-1}(H)|0\rangle$, $T_{n-2}(H)|0\rangle$.
The moment $\mu_n$ is finally obtained from the scalar product
$\langle 0|T_n(H)|0\rangle$ according to Eq.~\eqref{ChebyImpurity1}.
During the iteration the index of any vector $|n\rangle$ is increased
at most by one in Eq.~\eqref{HamMatrix},
and the vector $T_n(H)|0\rangle$ is a linear combination of Chebyshev
vectors $|m\rangle$, with $m \le n$.
The $n^{th}$ Chebyshev moment $\mu_n$ of $A(\omega)$ is therefore
obtained from the first $n$ Chebyshev moments
$\mu_0^B,\dots,\mu_n^B$ of $A_B(\omega)$.
We have thus devised a computational scheme to map $n$
moments of a given spectral function $A_B(\omega)$ to one part $H_B$
of $H$ to $n$ moments of a spectral function $A(\omega)$ to the full
Hamiltonian $H$ \textit{without} resorting to an explicit
representation~\eqref{HamBath} of $H_B$ in an orbital basis
$f^\dagger_\alpha$. This is the essence of CSM.

Note that only spectral functions $A_B(\omega)$, $A(\omega)$ occur in the
calculation.
The `missing' real part of the corresponding Green function, which is
needed in Eq.~\eqref{Gexact}, is implicitly
accounted for by causality relations preserved in the calculation.
As shown in  Ref.~\onlinecite{WWAF06}, we can obtain the Green
function $G(\omega)$ from moments $\mu_n$ of the spectral function
$A(\omega)$ without invoking Kramers-Kronig-relations.

\subsection{Scaling of $H$ and $H_B$}\label{Sec:Scaling}

We have so far omitted the scaling of $H$ and $H_B$ to
the interval $[-1,1]$.
To determine the scaling of $H_B$ 
we must choose a scaling interval $I_B$
that contains the domain of non-zero values of $A_B(\omega)$,
i.e. $A_B(\omega) = 0$ for $\omega \not\in I_B$ (cf. Sec~\ref{Sec:Cheby}).
For 
$I_B=[\omega^B_\mathrm{min},\omega^B_\mathrm{max}]$, 
the scaling factors
\begin{equation}
r=(\omega^B_\mathrm{max}-\omega^B_\mathrm{min})/2 \;, \; 
s=(\omega^B_\mathrm{min}+\omega^B_\mathrm{max})/2
\end{equation}
give the scaling $H_B = r \tilde{H}_B+s$ of $H_B$.
The smaller $I_B$, the higher is the resolution of
$A_B(\omega)$ for a given number of moments.
Similarly, we choose a scaling interval $I$ with
$A(\omega)=0$ for $\omega \not\in I$.
With scaling factors $p,q$ determined from $I$ as before, we find
for the scaling of $H$
\begin{equation} \label{HScaling}
\tilde{H} = \frac{1}{p}(H-q) = \frac{-\Delta}{p} d^\dagger d +
\frac{r}{p} \tilde{H}_B + \frac{s-q}{p} \;.
\end{equation}
It is straightforward to introduce the scaling factors into
Eqs.~\eqref{ChebyImpurity2},~\eqref{ChebyImpurity3},
and the matrix form~\eqref{HamMatrix}.

Since we do not know $A(\omega)$ in advance, we have to rely on estimates
for $I$.
For Eq.~\eqref{HOneSite},
the term $-\Delta d^\dagger d$ has eigenvalues $0$ and $-\Delta$.
It follows that any $I$ with $I \supset I_B \cup (I_B -\Delta)$ is a
possible choice.
Similar estimates can be obtained in other situations.
Alternatively we can determine the minimal and maximal eigenvalue of
$H$ by the Lanczos algorithm (cf. Sec.~\ref{Sec:Bosonic}).
Note that precise knowledge of the minimal or maximal eigenvalue
of $H$ is not necessary, but any bound works. 
It often suffices to obtain an estimate from operator
norms.
A factor of the order of two in the scaling is
tolerable for most practical purposes.
The resolution of CSM -- or KPM -- still compares
quite favorably to the resolution of other techniques.
The possible loss of resolution for a too large $I$ can be 
compensated for by increasing the number of moments in the
calculation.
We show in Sec.~\ref{Sec:TruncatedHB}, that we can even increase the
resolution for $A(\omega)$ without increasing the number of moments for
$A_B(\omega)$.

The reader should be reminded that
the need to consider scaling is the price one has to pay for the high
resolution of the Chebyshev technique -- this is  certainly
no drawback of the method.
Interestingly, there are situations where 
the divergence of moments for wrong scaling can be used to our advantage:
We will exploit it below to determine the
groundstate energy of $H$.

Now consider the case $\Delta=0$ in Eq.~\eqref{HOneSite}.
We know that then $A(\omega)=A_B(\omega)$.
With scaling intervals $I \supset I_B$,
the moments $\mu_n$, $\mu^B_n$
of these two identical functions are different since $I \ne I_B$.
We can use CSM to calculate the $\mu_n$, reproducing 
$A_B(\omega)$ on the larger interval $I$ (see next subsection).
Now assume that we choose $I_B$ much larger than necessary. 
In principle, a possible scaling interval $I$ for $H$ 
might then be smaller than $I_B$, i.e. $I \subset I_B$.
Will CSM then reproduce the spectral function
$A_B(\omega)$, respectively its moments, on the smaller
interval $I$?

The answer is no: The interval $I$ must always be larger than $I_B$,
i.e. $I \supset I_B$.
Then, and only then, the Chebyshev iteration for $\Delta=0$,
and $H=H_B$, is stable in the space $\mathcal{H}_c$.
This can be understood from the fact that 
$H_B$ acts within $\mathcal{H}_c$ like an operator with unity norm:
The column sums of the matrix~\eqref{HamMatrix} for $\Delta=0$ are
unity, and the first column has a single unity entry.
Note that this is a statement about the norm of $H_B$ as a matrix, and
not with respect to the scalar product of Chebyshev vectors.
If now $I \subsetneqq I_B$, then $r/p>1$, and $(r/p) \|\tilde{H}\|>1$ in Eq.~\eqref{HScaling}.
Then,
the coefficients in front of the Chebyshev vectors $|n\rangle$ diverge
during the iteration, 
while the moments $\mu_n^B$ decay with the same rate.
Each moment $\mu_n$ is a sum of products of coefficients and moments
$\mu_n^B$.
Within exact arithmetics the correct moments $\mu_n$ would be reproduced.
In finite precision arithmetics however, numerical round-off errors
entirely ruin the output.
On the contrary, for $I \supset I_B$, the recursion is perfectly
stable, up to any number of moments. This is an important point to be aware of.

\subsection{Numerical results}\label{Sec:NumRes}

\begin{figure}
\includegraphics[width=0.5\textwidth]{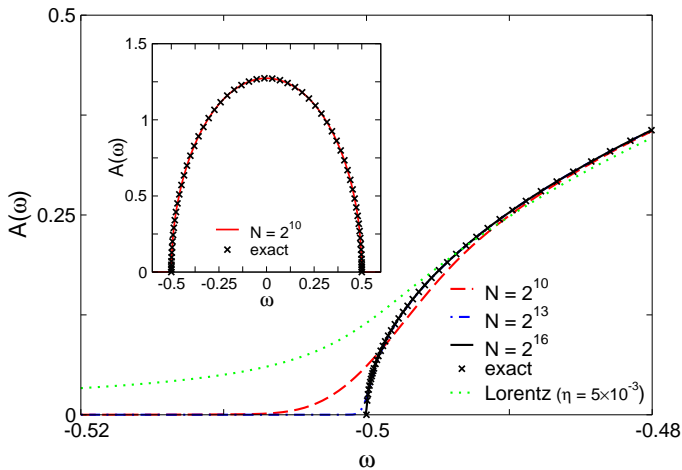}
\caption{(Color online) Spectral function $A(\omega)$ for semi-circular
  $A_B(\omega)$ according to Eq.~\eqref{ASemiCirc},
  with $W=1$ and $\Delta=0$.
  The scaling intervals are $I_B=[-0.6 W,0.6 W]$ and $I=[-1.2 W, 1.2 W]$,
  i.e. $r/p = 0.5$ in Eq.~\eqref{HScaling}.
  The figure 
  displays the region around the lower band edge $\omega=-W/2$,
  with curves for $N=2^{10},2^{13},2^{16}$ Chebyshev moments,
  the exact result, and a curve with Lorentzian broadening 
  $\eta/(\omega^2+\eta^2)$ to $\eta=5 \times 10^{-3}$.
  The latter curve illustrates the typical quality of results from
  the Lanczos Recursion Method.
  With $N=2^{10}$ Chebyshev moments the
  exact curve is much better reproduced than with Lorentzian broadening.
  The inset shows the full curve, whose deviation from the exact result is
  below linewidth on this scale.
  The curve for $N=2^{16}$ lies on top of the exact result 
  even in magnification of the band edge,
  and demonstrates that CSM can achieve arbitrary precision.}
\label{Fig:Impurity1}
\end{figure}

We now show results from CSM for the model defined in Eq.~\eqref{HOneSite}.
We start with a semi-circular spectral function
\begin{equation}\label{ASemiCirc}
  A_B(\omega) = \frac{8}{\pi W^2}\sqrt{W^2/4-\omega^2} \;,
\quad |\omega|\le W/2 \;,
\end{equation}
realized e.g. as the density of states in a Bethe lattice.
$W$ is the band width.
The first test of CSM 
is the calculation of $A(\omega)$ for $\Delta=0$ 
in Fig.~\ref{Fig:Impurity1}.
Then, $A(\omega)=A_B(\omega)$, and the numerics has to reproduce
Eq.~\eqref{ASemiCirc} on an interval $I$ different from $I_B$.
In particular, the moments $\mu_n$ and $\mu^B_n$ are not identical.

We expect strongest deviations from the numerical result to
Eq.~\eqref{ASemiCirc} close to the band edges, where the spectral
function behaves like a square root.
For $N=2^{10}$ moments, the absolute deviation 
$|A^n(\omega)-A^e(\omega)|$ between numerical ($A^n(\omega)$) and
exact ($A^e(\omega)$) spectral function 
is smaller than $6 \times 10^{-2}$ at the band edge,
and smaller than $5 \times 10^{-5}$ over the remaining $90 \%$ of the band.
The cumulative deviation $\int |A^n(\omega)-A^e(\omega) | d\omega$ is
below $10^{-4}$.
Notably, the resolution is much higher than for other methods,
e.g. using the Lanczos Recursion Method with
  Lorentzian broadening of the spectral function. 
Note also that the two-fold Chebyshev recursion -- once for the
moments of $A_B(\omega)$, once for $A(\omega)$ -- is perfectly stable
for any number of moments.
The calculation of many thousand of Chebyshev moments is no problem,
allowing for unprecedented resolution.

\begin{figure}
  \includegraphics[width=0.5\textwidth]{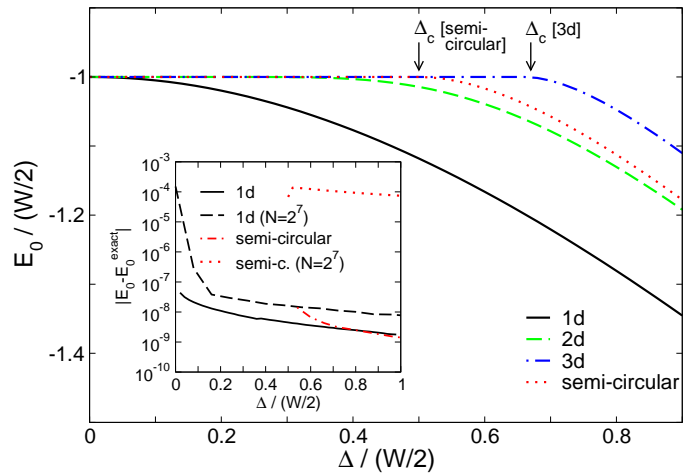}
  \caption{(Color online) Groundstate energy $E_0(\Delta)$ 
    for an impurity in a 1d, 2d, 3d lattice, and with a
    semi-circular density of states.
    $E_0(\Delta)$ is calculated with the method described in text,
    for $N=2^{10}$ Chebyshev moments.
    The arrows indicate the critical $\Delta_c$,
    with $\Delta_c[\text{semi-circular}] = W/4$ and
    $\Delta_c[3d] \approx 0.330 \times W$.
    The inset displays the error to the exact result for 1d and
    semi-circular density of states,
    where a simple exact result for $E_0(\Delta)$ is available.
    Already for $N=2^7$ moments we obtain results with an error below
    $10^{-4}$.
    For 1d, the error decreases rapidly for larger $\Delta$, since the
    groundstate is then strongly localized at the impurity, and the band
    edge singularity of the density of states less important.
  }
  \label{Fig:Impurity2}
\end{figure}

\begin{figure}
\includegraphics[width=0.5\textwidth]{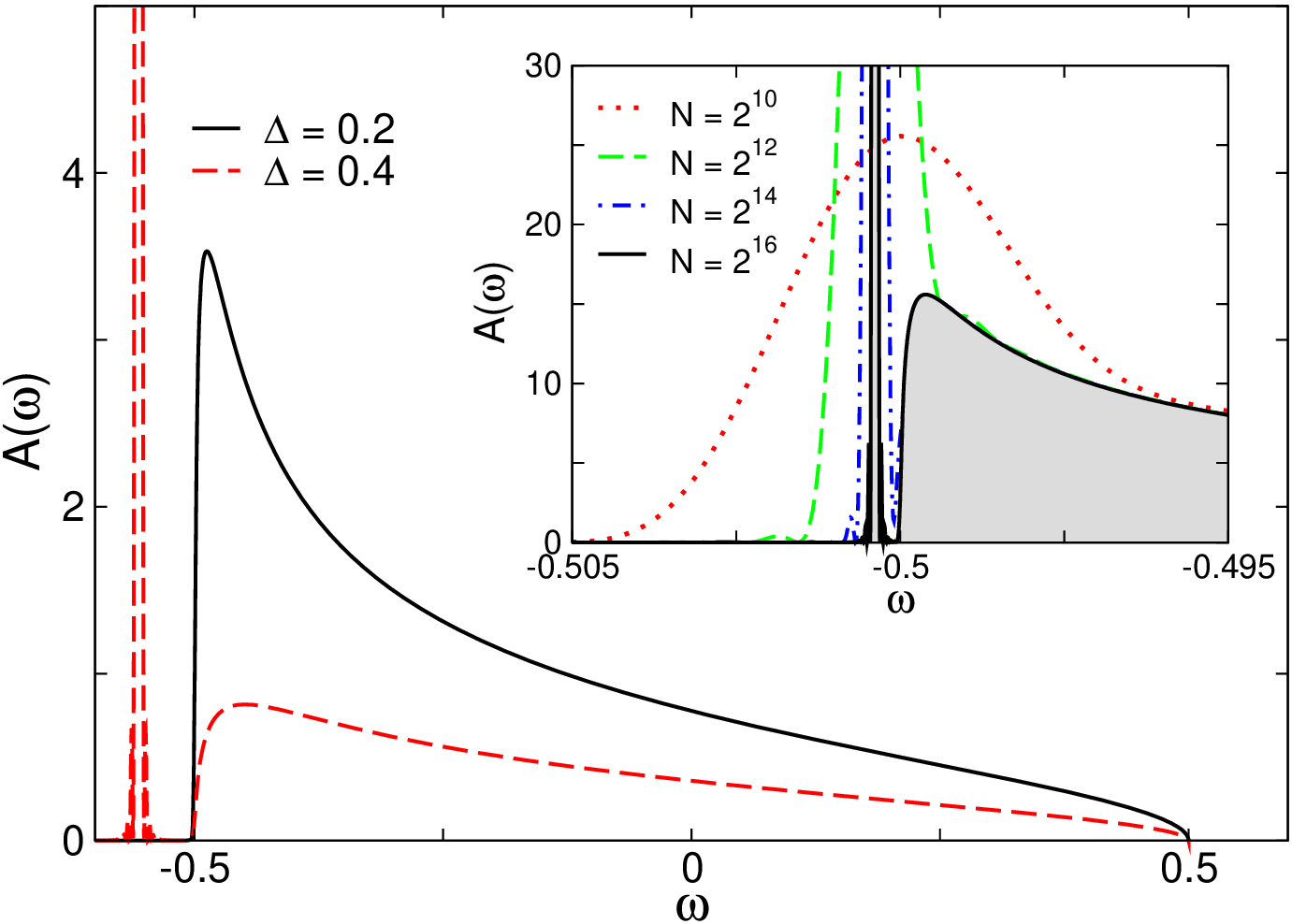}
\caption{(Color online) Spectral function $A(\omega)$ for semi-circular
  $A_B(\omega)$ (Eq.~\eqref{ASemiCirc}),
  with $W=1$ and finite $\Delta$.
  The scaling intervals are $I_B=[-(0.5+\epsilon) W, (0.5+\epsilon)
  W]$ and $I=[-(0.5+\epsilon)W-\Delta,(0.5+\epsilon) W]$,
  with a small offset $\epsilon=10^{-4}$.
  For $\Delta > \Delta_c=W/4$, $A(\omega)$ has a pole outside the band
  of continuum states (cf. Fig.~\ref{Fig:Impurity2}).
  The deviation of the curves with $N=2^{10}$ to the exact result
  is below linewidth, except for the pole in the curve for
  $\Delta=0.4$, which attains a finite width.
  The inset displays four curves for $\Delta=0.26$ and different $N$.
  For this $\Delta$, $A(\omega)$ has a pole at $E_0(\Delta) \approx
  -W/2-3.846\times 10^{-4}$.
  Although the pole is separated from the band edge $-W/2$ by less than
  $10^{-3}$ times the bandwidth $W$,
  it can be resolved within CSM provided $N$ is large enough.
  It is hardly possible to achieve a similar increase in resolution
  with e.g. the Lanczos Recursion Method.
  }
\label{Fig:Impurity3}
\end{figure}

To further test the accuracy of our approach
we calculate the groundstate energy $E_0(\Delta)$ of an impurity in
an infinite chain (1d), square (2d), cubic (3d) tight-binding lattice,
and for the semi-circular density of states in Eq.~\eqref{ASemiCirc}
(see Fig.~\ref{Fig:Impurity2}).
For sufficiently large $\Delta$
a bound state at the impurity exists, with energy outside the band of
continuum states, i.e. $E_0(\Delta) < -W/2$.
In 1d and 2d, a bound states exists for any $\Delta>0$, while in
3d and for the semi-circular density of states a bound state exists
above a critical value $\Delta_c$.
In Fig.~\ref{Fig:Impurity3} we show the impurity spectral
function for $\Delta \ne 0$.
$A(\omega)$ has a $\delta$-peak at $\omega=E_0$ if $\Delta > \Delta_c$.
Even if the peak is very close to the band edge it can be resolved
within CSM by increasing the number of moments $N$.

To find $E_0(\Delta)$ within CSM, we exploit the possible divergence
of moments in the Chebyshev recursion.
We initially choose a scaling interval
$I=[\omega_\mathrm{min},\omega_\mathrm{max}]$ for $H$
so large that moments do not diverge.
Then we vary $\omega_\mathrm{min}$.
If the moments diverge, $E_0 < \omega_\mathrm{min}$, otherwise $E_0 \ge
\omega_\mathrm{min}$.
Using e.g. a bisection algorithm on $\omega_\mathrm{min}$ we
can calculate $E_0$ to high precision.
The computational effort for Fig.~\ref{Fig:Impurity2} is independent
of the lattice dimension, as only a fixed number of moments enter the
calculation. The most demanding case here is indeed 1d, as the
spectral function diverges at the band edge.
Despite this particular complication, the results are extremely accurate 
already for $N=2^{10}$ moments.

For our example, $E_0$ is the groundstate energy of a single particle,
i.e. the smallest energy with $A(E_0) \ne 0$.
The weight $A(E_0)$ can be arbitrarily small.
From Eq.~\eqref{Gexact}, we see that $A(\omega) \ne 0$ whenever
$A_B(\omega) \ne 0$.
Within the numerics the domain of non-zero values of
$A_B(\omega)$ is fixed by the scaling interval
$I_B=[\omega^B_\mathrm{min},\omega^B_\mathrm{max}]$.
We also know from the previous subsection, that
moments diverge for $\omega_\mathrm{min} > \omega^B_\mathrm{min}$.
As a consequence, the calculation results in 
$E_0 \le \omega^B_\mathrm{min}$.
Therefore $\omega^B_\mathrm{min}$  must be chosen at the lower band edge
of $A_B(\omega)$, i.e. $\omega^B_\mathrm{min}=-W/2$ in all examples.
As long as $|E_0| > |\omega^B_\mathrm{min}|$, 
especially for $\Delta > \Delta_c$, no problem occurs.
Note that the situation is different at finite fermion density,
where the groundstate energy is determined by filling states up to a
certain density $n$, with $n=\int_{-\infty}^{E_0} A(\omega) d\omega$.
Negligible weight in $A(\omega)$ does not change $E_0$ then,
and the calculated $E_0$ does not depend that strongly on $I_B$.

From the examples shown here we conclude that CSM 
provides highest accuracy.
The computational effort to obtain the results shown is small --
a calculation for $N=2^{10}$ takes less than a second --
which leaves plenty of room for applications to less simple problems.
In the next paragraph we discuss how 
the computational effort can be further reduced,
by adjusting the resolution of the two Chebyshev expansions for
$A(\omega)$ and $A_B(\omega)$.

\section{Properties of the truncated Chebyshev representation of $H_B$}\label{Sec:TruncatedHB}

In the examples of the previous section we calculate $N$ moments of
$A(\omega)$ from the same number of moments of $A_B(\omega)$.
Calculations can also be performed for a different number of moments
of $A_B(\omega)$, say $M$.
For $M\ge N$, the moments of $A(\omega)$ do not change
and the calculation is exact, as we noted above.
For $M < N$ however, the calculation is approximate and its result depends on $M$.

The computational effort for given $M,N$ scales as $O(MN)$ for time,
and $O(M)$ for memory: Calculating more moments of $A(\omega)$ requires only
more time, while using more moments of $A_B(\omega)$ requires more time and
memory.
It is therefore natural to ask whether we can obtain
$N$ moments of $A(\omega)$ from fewer moments of $A_B(\omega)$,
especially if we remember that both are calculated with different
scaling for $H$ and $H_B$.
We know that the resolution of a Chebyshev expansion increases
linearly with the number of moments,
as can be deduced e.g. from the width (standard deviation) $\sigma =
\pi/N$ of the Jackson kernel, see Appendix~\ref{App:Cheby}.
If we demand that the resolution for  $A(\omega)$ and
$A_B(\omega)$ is the same, we obtain the condition $M/N \gtrsim r/p$,
with the scaling $H = p \tilde{H} +q$, $H_B = r \tilde{H}_B +s$ as
in Eq.~\eqref{HScaling}.
As we discussed there, the factor $r/p$ is always smaller than unity.
For the example of the Holstein model in Sec.~\ref{Sec:Holstein}, it
is easily of the order $10^{-1}$.

\begin{figure}
\includegraphics[width=0.5\textwidth]{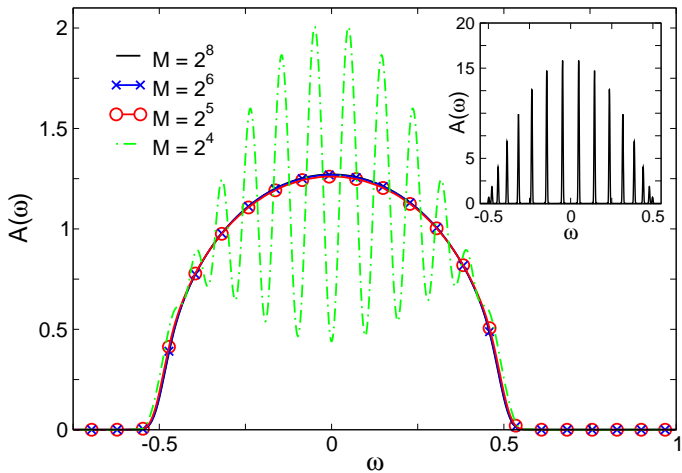}
\caption{(Color online) 
  Similar to Fig.~\ref{Fig:Impurity1},
  this figure shows the spectral function $A(\omega)$ for semi-circular
  $A_B(\omega)$ according to Eq.~\eqref{ASemiCirc},
  with $W=1$ and $\Delta=0$.
  The scaling intervals are
  $I_B=[- (0.5+\epsilon) W, (0.5+\epsilon) W]$
  with $\epsilon=10^{-4}$ and $I=[-2W,2W]$,
  i.e. $r/p = 1/4$ in Eq.~\eqref{HScaling}.
  The calculation is performed for $N=2^8$ Chebyshev moments of
  $A(\omega)$ and $M$ moments of $A_B(\omega)$.
  As long as $M/N \gtrsim r/p$, i.e. here $M \gtrsim 2^6$,
  the result can not be distinguished from the (numerically exact)
  curve for $M=N$.
  For $M=2^5$, deviations occur at the band
  edges, and for $M=2^4$ single spurious peaks are resolved.
  The inset displays a calculation for $N=2^{11}$ and $M=2^4$.
  Even for this strong mismatch, a perfectly positive 
  spectral function is obtained,
  resolving $\delta$-peaks at $\omega_j = r x_j + s$
  corresponding to the roots $x_j$ of $T_M(x)$
  (see text).}
\label{Fig:Scaling}
\end{figure}

In Fig.~\ref{Fig:Scaling} 
we show $A(\omega)$ for different ratios $M/N$.
As long as the condition $M/N \gtrsim r/p$ is
fulfilled, we produce an absolutely accurate $A(\omega)$ from $M<N$
moments of $A_B(\omega)$, while the computational effort is
considerably reduced.
In the worst case, for $M/N \ll r/p$, discrete energy
levels in the truncated bath Hamiltonian $H^M_B$ are resolved.
Even $N\gg M$ does not lead to erroneous results as a negative
spectral function.
In particular, the CSM is perfectly stable for
arbitrary $N$ and $M$.

Using $M < N$ moments of $A_B(\omega)$ in the calculation is
equivalent to working with a truncated bath Hamiltonian $H^M_B$ on the
$M$-dimensional subspace $\mathcal{H}^M_c$ of $\mathcal{H}_c$ that
is spanned by vectors $|0\rangle, \dots, |M-1\rangle$.
The matrix of $H^M_B$, with $H^M_B|n\rangle = \sum_m {(H^M_B)}_{mn}
|m\rangle$, is according to Eq.~\eqref{ChebyImpurity2} the tridiagonal
$M\times M$-matrix 
\begin{equation}\label{HBTruncated}
  (H^M_B)_{mn} = \frac{1}{2} \begin{pmatrix} 0 & 1 \\ 2 & 0 & 1 & & \\
    & 1 & 0 & 1 \\ & & 1 & 0 & \smash{\ddots} \\ & & & \ddots & 0 & 1 & \\
& & & & 1 & 0 & 1 \\
& &  & & & 1 & 0 
  \end{pmatrix} \;.
\end{equation}

We show in App.~\ref{App:Recurrence}, that the characteristic
polynomial of $(H^M_B)_{mn}$ is 
\begin{equation}
\det[x-(H^M_B)_{mn}] = 2^{-(M-1)} T_M(x) \;.
\end{equation}
The $M^{th}$ Chebyshev polynomial $T_M(x)=\cos(M \arccos x)$
has $M$ distinct real roots $x_j= \cos \frac{\pi (j-1/2)}{M}$ for
$j=1,\dots,M$.
It follows that $(H^M_B)_{mn}$ has $M$ distinct
real eigenvalues, hence $H^M_B$ is a diagonalizable $M\times
M$-matrix with real eigenvalues $x_j$.
The position of the discrete energy levels resolved for $N \gg M$  in
Fig.~\ref{Fig:Scaling} is determined by the $x_j$.
We discuss in App.~\ref{App:Eigen} how the weight of the corresponding
peaks can be obtained from the moments $\mu^B_m$.

\section{Linear chain coupled to a bath}
\label{Sec:Chain}

A different example is provided by an electron hopping along a finite
chain of length $L$ that is connected to a bath at its `right' end
(site $L$).
The Hamiltonian 
\begin{equation}\label{HamChain}
  H = -t \sum_{i=1}^{L-1} (c^\dagger_{i+1} c_i  + c^\dagger_i c_{i+1} )  \;
+ H_L + H_B
\end{equation}
has the form of Eq.~\eqref{HamGeneral}.
Here, $c^\dagger_i$ creates an electron at lattice site $i$,
and the coupling between chain and bath is $H_L=-t (d^\dagger c_L +
c^\dagger_L d)$, as in Eq.~\eqref{HamLink2}, with $V=-t$.
That the bath couples to the $d$-orbital implies that
$[c^{(\dagger)}_i,H_B]=0$, but $[d^{(\dagger)},H_B] \ne 0$.

In difference to the example in Sec.~\ref{Sec:Host} (see
Eq.~\eqref{HOneSite}) the particle hops to and from the bath, hence
the Hilbert space $\mathcal{H}=\mathcal{H}_S \oplus \mathcal{H}_c$ of
the problem is the direct sum of the
Hilbert space $\mathcal{H}_S$ of the chain and the CS $\mathcal{H}_c$.
A basis of $\mathcal{H}_S$ 
consists of vectors $|\psi_i\rangle = c^\dagger_i |\mathrm{vac}\rangle$,
for the electron at site $i$.
The vectors $|\psi_i\rangle$ are orthogonal to the Chebyshev vectors
$|n\rangle$.
The operation of $H$ is summarized in
\begin{equation}
\begin{split}
  H |\psi_1\rangle &= -t |\psi_2\rangle \;,\\ 
  H |\psi_i\rangle &= -t |\psi_{i+1}\rangle \, -t|\psi_{i-1}\rangle \\
  & \qquad \text{ for } i=2,\dots,L-1 \; , \\
  H |\psi_L\rangle &= -t |0\rangle -t|\psi_{L-1}\rangle \;, \\
  H |n\rangle &= -t \mu^B_n |\psi_L\rangle + H_B |n\rangle \;,
\end{split}
\end{equation}
with the missing equations supplied by Eq.~\eqref{ChebyImpurity2}.

The calculation
of the spectral function 
$A_{11}(\omega) = \langle\mathrm{vac}|c_1 \delta(\omega-H)
c^\dagger_1|\mathrm{vac}\rangle$ -- the spectral function to the
`left' end of the chain in Fig.~\ref{Fig:Bath} --
to given bath spectral function $A_B(\omega)$
proceeds along the lines established in Sec.~\ref{Sec:Host}.
In contrast to the example~\eqref{HOneSite}, where
Eq.~\eqref{ChebyImpurity1} was used, scalar products with
the starting vector $c^\dagger_1|\mathrm{vac}\rangle$ 
of the Chebyshev iteration do not involve the $\mu^B_n$.

\begin{figure}
\includegraphics[width=0.5\textwidth]{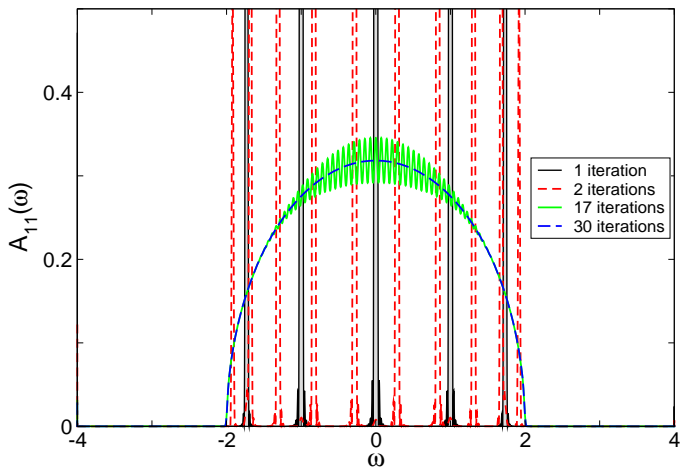}
\caption{(Color online) 
Spectral function $A_{11}(\omega)$ for a chain with $L=5$ sites, calculated
for $N=N_B=2^{10}$ moments.
In the first iteration, $A_{11}(\omega)$ consists of $L$ peaks.
The number of peaks is increased by $L$ per iteration (see curve for 2
iterations).
Peaks merge if their distance is smaller than the
resolution (see curve for $17$ iterations), until $A_{11}(\omega)$
is converged to the self-consistent semi-circular solution after a
sufficient number of iterations (here about $30$ iterations).
The scaling intervals are
$I_B=[-(W/2+\epsilon),(W/2+\epsilon)]$,
with $\epsilon=10^{-4}$, and $I=[-W,W]$, with $W=4t$.
Note that the number of iterations until convergence depends on the
number of moments, i.e. the resolution.
For eg. $N=2^7$ moments, convergence is obtained after $6$ iterations.}
\label{Fig:Chain}
\end{figure}

To make the present example self-consistent we demand that 
$A_{11}(\omega) = A_B(\omega)$.
The self-consistent solution $A_{11}(\omega)$ is the
spectral function of a half-infinite chain at its open end, i.e. the
semi-circular spectral function from Eq.~\eqref{ASemiCirc} with $W=4t$.
To obtain the self-consistent $A_{11}(\omega)$ we 
start with an initial guess for $A_B(\omega)$, e.g.
setting $\mu^B_n=0$ corresponding
 to $A_B(\omega)=0$, and calculate
the moments $\mu_n$ of $A_{11}(\omega)$.
Note that the choice $\mu^B_n=0$ does not correspond to a spectral
function, as the sum rule $\int A_B(\omega) d\omega = \mu^B_0 = 1$ is
violated. This does not matter for the calculation, and we could
obtain the same effect by setting $V=0$.
We start a new calculation taking the $\mu_n$ just calculated
as new bath moments $\mu^B_n$,
and reiterate this calculation until the moments $\mu_n$, or
equivalently $A_{11}(\omega)$, are converged (see
Fig.~\ref{Fig:Chain}).

We found in Sec.~\ref{Sec:Host} that we must choose the
scaling interval $I$ for $A_{11}(\omega)$ larger than $I_B$ for
$A_B(\omega)$.
If we start a new iteration with the previously calculated moments $\mu_n$
replacing $\mu^B_n$
we must also replace the interval $I_B$ by $I$, and consequently allow
for growing scaling intervals -- and corresponding loss
of resolution -- during the iterations.
As an alternative, we keep the interval $I_B$ fixed and 
rescale $A_{11}(\omega)$ from $I$ to $I_B$ in our implementation.
We first construct $A_{11}(\omega)$ on the interval $I$ from the
$\mu_n$, then rescale it in $\omega$-space to the interval $I_B$
with a linear transformation $\omega \mapsto (r/p)(\omega-q) +
s$ (cf. Eq.~\ref{HScaling}),
and finally feed the moments of this rescaled spectral function as new
$\mu^B_n$ back into the calculation.
In the rescaling we can check whether we throw away significant weight
of $A_{11}(\omega)$ which signals a too small $I_B$.
For some examples where we do not
know the relevant interval $I_B$ in advance, 
e.g. for the Holstein polaron (Sec.~\ref{Sec:Holstein}), we let the program
determine a suitable $I_B$ that contains all but negligible weight of
$A_{11}(\omega)$.
Typically, the interval changes during the first few iterations and
then stabilizes with convergence of $A_{11}(\omega)$.
The rescaling transformation needs two Fourier transforms (preferentially
FFT) for the calculation of the spectral function from the moments and
vice versa, and one interpolation for the linear scaling.
We use spline interpolation, but for not too few moments everything also works
without interpolation.

The reader should note that the rescaling transformation -- which
shrinks the $\omega$-interval of the Chebyshev expansion -- is a linear
transformation of moments $\{\mu_n\} \mapsto \{\mu^B_m\}$.
A calculated $\mu^B_m$ depends on every $\mu_n$.
In contrast, the CSM of Sec.~\ref{Sec:Host} implements
for $\Delta=0$ in Eq.~\eqref{HOneSite}
a transformation $\{\mu^B_m\} \mapsto \{\mu_n\}$,
that blows up the $\omega$-interval (cf. Fig.~\ref{Fig:Impurity1}).
An inspection of Eqs.~\eqref{ChebyImpurity1},~\eqref{ChebyImpurity2}
shows that this transformation is also linear. 
For $\Delta \ne 0$, it ceases to be linear as the $\mu^B_n$ also occur
in $H$ itself (cf. Eq.~\eqref{HamMatrix}),
by virtue of Eq.~\eqref{ChebyImpurity4}.
The transformation has the property that each calculated $\mu_n$ depends
only on $\mu^B_m$ with $m\le n$.
We now understand a second reason why the Chebyshev iteration has to be
unstable for $I_B \not\subset I$.
If the Chebyshev iteration could be used to shrink the
$\omega$-interval -- that is for $I \subset I_B$ -- every calculated
moment ($\mu_n$) had to depend on every moment supplied to the
calculation ($\mu^B_n$), in contradiction to the properties of the recursion.

We discussed in Sec.~\ref{Sec:TruncatedHB} how to
calculate with CSM $N$ moments $\mu_n$ from $M$ moments
$\mu^B_m$ for $M<N$, depending on the scaling intervals
$I$ and $I_B$.
Do we waste part of the numerical results 
if again only $M<N$ moments $\mu^B_m$ are calculated in the rescaling
transformation? 
Obviously not: For a smaller interval less moments are
needed to obtain the same resolution.
In both directions, via CSM or the rescaling
transformation, the number of moments should be related by the
estimate $M/N \sim r/p$.

\section{Concurrent dynamics}\label{Sec:Concurrent}

The general Hamiltonian Eq.~\eqref{HamGeneral}
operates on a product space $\mathcal{H}=\mathcal{H}_S \otimes
\mathcal{H}_B$.
$H_S$ ($H_B$) operates on $\mathcal{H}_S$ ($\mathcal{H}_B$),
and $H_L$ links the two spaces.
In CSM, $\mathcal{H}_B = \mathcal{H}_c$.
We can generally write the Hilbert space of $H$ in such a way, with 
$\mathcal{H}_S$ and $\mathcal{H}_B$ as Fock spaces,
but additional constraints may single out a subspace.
In the example of the linear chain with the restriction to a single electron,
$\mathcal{H}$ is the direct sum of $\mathcal{H}_S$ and $\mathcal{H}_c$.
 
Assume that system and bath do not couple, i.e. $H_L=0$ and $H=H_S+H_B$.
As $H_S$ and $H_B$ commute,
the spectral function
$A(\omega)=\langle\psi|\delta(\omega-H)|\psi\rangle$ to a product
state $|\psi\rangle = |\psi_S\rangle \otimes |\psi_B\rangle $ is the
convolution of the spectral functions of $H_S$ and $H_B$.
Explicitly,
\begin{equation}\label{AConvolution}
  A(\omega) = \sum_\alpha |\langle \alpha|\psi_S\rangle|^2
  A_B(\omega-\epsilon_\alpha) \;,
\end{equation}
where the sum is over eigenstates $|\alpha\rangle$ with eigenvalue
$\epsilon_\alpha$ of $H_S$, 
and $A_B(\omega) = \langle\psi_B|\delta(\omega-H_B)|\psi_B\rangle$.
To evaluate this expression we must diagonalize $H_S$ in advance,
to obtain its eigenvalues and
eigenstates.
It is therefore difficult to use Eq.~\eqref{AConvolution}
for even moderately complicated $H_S$.
Within CSM we calculate $A(\omega)$ without
diagonalization of $H_S$.
The convolution in
Eq.~\eqref{AConvolution} is implicitly performed in course of the
Chebyshev iteration.
This feature is essential for all situations where system and bath
degrees of freedom evolve in parallel.

\begin{figure}
\includegraphics[width=0.5\textwidth]{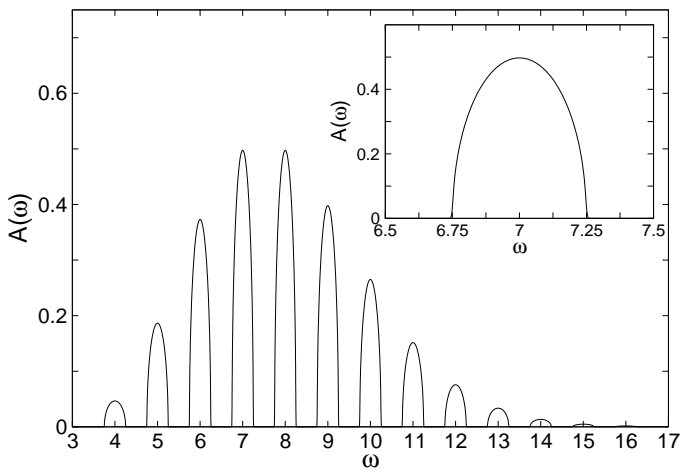}
\caption{Spectral function $A(\omega)$ from Eq.~\eqref{ASudden},
for $\omega_0=1$, $\epsilon_p=4$ in Eq.~\eqref{HIndBos},
and $W=0.5$ in Eq.~\eqref{ASemiCirc}.
We use $N=2^{13}$ moments and $M=2^8$ bath moments.
For maximally $n_b=25$ bosons in the calculation,
the scaling of $H$ and $H_B$ is $r/p \leq W/(n_b\omega_0)=0.02$.
According to Sec.~\ref{Sec:TruncatedHB},
it is sufficient to use about $M=160$ bath moments. 
The inset displays in magnification, how accurately the semi-circular
spectral function is resolved for a single subband.}
\label{Fig:BosonicSite}
\end{figure}

We want to illustrate this point
for the Hamiltonian
\begin{equation}\label{HIndBos}
H_S = -\sqrt{\epsilon_p \omega_0} (b^\dagger+b) c^\dagger c
+ \omega_0 b^\dagger b
\end{equation}
of a bosonic site, the independent boson model~\cite{Mah00}.
It is a simple model for electron-phonon coupling of localized
electrons,
if the bosons ($b^\dagger$) parametrize the elongation of an
ion that produces an electric field which
shifts the energy of an electron at the site ($c^\dagger$).
The groundstate of the bosonic part of $H_S$ in presence of a fermion, to energy $E_0=-\epsilon_p$, is the coherent state
\begin{equation}
|\mathrm{coh}\rangle = e^{-g^2/2} \sum_{n=0}^\infty \frac{g^n}{\sqrt{n!}}
(b^\dagger)^n  |\mathrm{vac}\rangle \;, \quad
g=(\epsilon_p/\omega_0)^{1/2} \;.
\end{equation}
For $A_B(\omega)$ we assume a semi-circular spectral function as in
Eq.~\eqref{ASemiCirc},
and prepare the system in the state $|\psi\rangle=|\mathrm{coh}\rangle
\otimes c^\dagger | \mathrm{vac}\rangle$.
We calculate the spectral function 
\begin{equation}\label{ASudden}
A(\omega)=\langle\psi| c^\dagger d \delta(\omega-H+E_0) d^\dagger c
|\psi\rangle \;.
\end{equation}
Physically, this corresponds to a sudden excitation of the electron
from the bosonic site ($c$-orbital) to a continuum of states given by $H_B$,
which is related to X-ray absorption of a localized electron~\cite{Mah00}.
We know that we should obtain $A(\omega)$ as a sum of semi-circular
spectral functions shifted by multiples of $\omega_0$, weighted with the
Poissonian distribution of bosons in $|\mathrm{coh}\rangle$.
The lowest band is centered at $\epsilon_p$, which is the energy to
remove the electron from the bosonic site.
In Fig.~\ref{Fig:BosonicSite} we show $A(\omega)$ calculated by CSM.
The numerical result perfectly agrees with the expected outcome. 

\section{A bosonic site coupled to a bath}\label{Sec:Bosonic}

We assumed in the previous section that $H_L=0$,
so that system and bath degrees of freedom do not mix.
Nothing changes to the applicability of our approach if this
restriction is abandoned, as the next examples show.
In contrast, the calculation of such `mixed' dynamics
is the intended application of CSM.

Let us combine the example from the previous section
with the example from Sec.~\ref{Sec:Host}.
We get the Hamiltonian 
\begin{equation} \label{HamHolOne}
  \begin{split}
  H  = &-\Delta c^\dagger c - \sqrt{\epsilon_p \omega_0} (b^\dagger + b)
  c^\dagger c + \omega_0 b^\dagger b \\
  &- t  (d^\dagger  c + c^\dagger d  ) + H_\mathrm{B}
\end{split}
\end{equation}
of a bosonic impurity ($c$-orbital) coupled to
a fermionic bath, e.g. a lattice.
Note that the impurity site is coupled to the bath via a term 
$H_L = - t  (d^\dagger  c + c^\dagger d  )$ as in
Eq.~\eqref{HamGeneral} or Eq.~\eqref{HamChain}.
We could also study a model with an bosonic impurity embedded in a
host similar to Eq.~\eqref{HOneSite},
but the present form is convenient for the study of the
Holstein model (Sec.~\ref{Sec:Holstein}).

For $\Delta <0$, the impurity is repulsive and acts as a static
barrier for electron motion.
Two competing mechanism determine the groundstate:
On the one hand, the energy of the impurity state is increased by $-\Delta$,
favoring a delocalized groundstate.
On the other hand, the formation of a localized polaron at the
bosonic impurity lowers the energy of the electron roughly by $\epsilon_p$.
A localized impurity state occurs if the loss in kinetic energy is
overcome by the gain in potential energy.
We know from Sec.~\ref{Sec:Host} that this happens -- for an
attractive impurity -- if $\Delta$ is larger than a critical
$\Delta_c$.
By a rough estimate, we expect here a localized groundstate for
$\epsilon_p \gtrsim \Delta_c-\Delta$.

To address this issue we must calculate the groundstate of the
model~\eqref{HamHolOne}.
In Sec.~\ref{Sec:NumRes} we explained how to determined the
groundstate energy $E_0$ using Chebyshev expansions,
testing for the divergence of Chebyshev moments for different scaling of $H$.
This time we must obtain the groundstate itself, not just its energy,
to be able to calculate correlation functions for the degrees of
freedom of the bosonic site.
To calculate the groundstate we use the Lanczos algorithm.
This requires our ability to perform
two operations: To apply $H$ to a vector, and to
calculate the scalar product between vectors.
For $H_B$ defined on the CS $\mathcal{H}_c$,
Eq.~\eqref{ChebyImpurity2} defines the application of $H_B$ to a
vector, and Eq.~\eqref{ChebyImpurity1} (or Eq.~\eqref{ChebyScal} in
App.~\ref{App:Herm}) gives the scalar product.
Note that for a successful application of Lanczos the moments $\mu^B_n$
should be modified by attenuation factors (Eq.~\eqref{JacksonFactors}),
cf. App.~\ref{App:Cheby} and~\ref{App:Herm}.

\begin{figure}
  \includegraphics[width=0.5\textwidth]{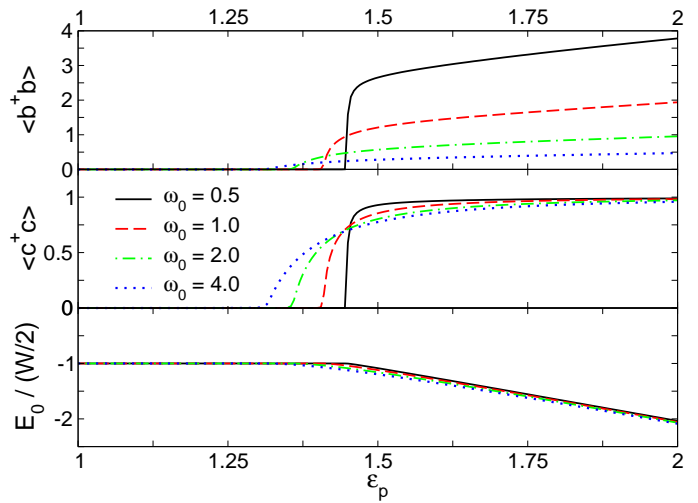}
\caption{(Color online)
  Average number of bosons $\langle b^\dagger b\rangle$,
  occupation probability of the impurity site $\langle c^\dagger c \rangle$,
  and groundstate energy $E_0$ 
  in dependence on $\epsilon_p$,
  for the model~\eqref{HamHolOne} with $\Delta / W=-1$
  and a semi-circular $A_B(\omega)$ with $W=1$ according to
  Eq.~\eqref{ASemiCirc}.
  Calculations have been performed with Lanczos/CS for $M=2^9$
  bath moments.
  Already $M=2^6$ moments produce accurate results away from the phase
  transition. Close to the phase transition, we can increase the
  accuracy easily by increasing $M$.
  Note that $M$ does not need to exceed the number of Lanczos iterations.
}
\label{Fig:Bosonic}
\end{figure}

In Fig.~\ref{Fig:Bosonic}
we show the groundstate energy $E_0$,
the occupation probability $\langle c^\dagger c\rangle$ and
the average number of bosons $\langle b^\dagger b \rangle$
calculated with the Lanczos algorithm and the CS.
As in Sec.~\ref{Sec:NumRes} the computational effort is independent of
the dimension.
We used a semi-circular $A_B(\omega)$ here, for a 3d lattice the
results are qualitatively the same.
While $E_0$ is a smooth function of $\epsilon_p$,
$\langle c^\dagger c\rangle$ and $\langle b^\dagger b \rangle$
signal the phase transition from a delocalized electron to a localized
polaron at a critical $\epsilon^c_p$.
For $\omega_0 \to \infty$, $\epsilon^c_p$ converges to the value
$\Delta_c-\Delta$ of the simple estimate.
For $\omega_0 \to 0$, the phase transition becomes more pronounced,
as a precursor of the first-order transition in the adiabatic limit
$\omega_0=0$.
Note that in contrast to the (Holstein) polaron problem, this model
has a phase transition also for finite $\omega_0$.

It is perhaps surprising that the Lanczos
algorithm, which constructs an orthonormal basis, works in combination
with the CS construction, which is based on non-orthogonal
vectors.
For $H=H_B$ the Lanczos algorithm constructs an orthonormal basis of
$\mathcal{H}_c$ and is equivalent to
Gram-Schmidt orthonormalization of the Chebyshev vectors $|n\rangle$.
It is known that the Gram-Schmidt procedure is prone to instabilities,
namely loss of orthogonality.
The same is true for the Lanczos algorithm,
which is a problem for the calculation of spectral functions.
The calculation of the groundstate however does not require
to construct an orthonormal basis of $\mathcal{H}_c$, and loss of
orthogonality is not a severe problem.
For the model~\eqref{HOneSite} or~\eqref{HamHolOne}
we can force the Lanczos algorithm to fail,
if $\Delta=\epsilon_p=0$ and $\omega_\mathrm{min}^B \ll -W/2$ for the
lower bound $\omega_\mathrm{min}^B$ of $I_B$.
For these parameters, the groundstate is a linear combination of 
Chebyshev vectors $|n\rangle$ with energy $E_0 >
\omega_\mathrm{min}^B$, and the Lanczos algorithm will not converge.
Remember also the discussion in Sec.~\ref{Sec:NumRes} concerning the
calculation of $E_0(\Delta)$ for a single fermion,
where we noted that we must guarantee $E_0 < \omega_\mathrm{min}^B$.
In practice, we never encountered a problem with the Lanczos algorithm in
combination with the CS construction.
For Fig.~\eqref{Fig:Bosonic}, Lanczos converges fastest for large
$\epsilon_p$, when the groundstate is localized at the impurity site.

\subsection{Holstein model}\label{Sec:Holstein}

The solution of the model~\eqref{HamHolOne} is related
to the self-consistent solution of the Holstein model
within DMFT.
The Holstein model~\cite{Ho59b} is a standard model for electron-phonon-coupling.
Its Hamiltonian
\begin{equation} \label{HamHol}
 H = -t_{ij} \sum_{ij} c^\dagger_i c_j \,
    - \sqrt{\epsilon_p \omega_0} \sum_{i} (b^\dagger_i + b_i)
  c^\dagger_i c_i + \omega_0 \sum_{i} b^\dagger_i b_i
\end{equation}
contains a local coupling of the electron density ($c^\dagger_i c_i$)
to dispersionless optical phonons ($b^\dagger_i$), in addition to the
kinetic energy of electrons modelled by the hopping term, and the
kinetic energy of phonons with frequency $\omega_0$.
The polaron shift $\epsilon_p$ is the
groundstate energy of the model for $t_{ij}\equiv 0$
(cf. Sec.~\ref{Sec:Concurrent}).

The solution of the Holstein model, especially for
spectral properties, is still a demanding problem (for a recent review
see Ref.~\onlinecite{FT07}).
We successfully used Chebyshev expansions to obtain spectral functions
or the optical conductivity for finite systems~\cite{SWWAF05,fahw06}.
Here we use DMFT
to relate the Holstein model to the model~\eqref{HamHolOne} where a
coupling to a bath occurs.

Within DMFT, the ($k$-integrated) spectral function
\begin{equation}
A(\omega) =  
\langle\mathrm{vac}|c_i \delta(\omega-H) c^\dagger_i |\mathrm{vac}\rangle
\end{equation}
is obtained as the self-consistent solution of an impurity model with a single
interacting site.~\cite{GKKR96}
For the Holstein model~\eqref{HamHol}, the impurity model is just the
model of a single bosonic site~\eqref{HamHolOne} for $\Delta=0$.
If we assume a semi-circular density of states for the non-interacting
problem, i.e. $A(\omega)$ has the functional dependence of
Eq.~\eqref{ASemiCirc} for $\epsilon_p=0$,
self-consistency is established by $A(\omega)=A_B(\omega)$.
For a different non-interacting density of states this relation is more
complicated, but the difference is not relevant in our context.
We have shown in Sec.~\ref{Sec:Chain}
how a self-consistent calculation is performed.
The complicated part of the DMFT calculation, 
to obtain $A(\omega)$ for the Hamiltonian~\eqref{HamHolOne} to given
$A_B(\omega)$, is solved by an application of CSM.

\begin{figure}
\includegraphics[width=0.5\textwidth]{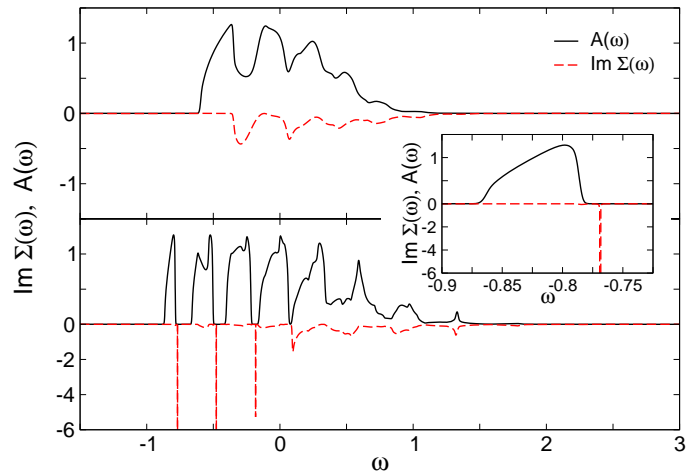}
\caption{(Color online) 
  Spectral function $A(\omega)$ and imaginary part of the selfenergy for
  the Holstein model~\eqref{HamHol} within DMFT, for a semi-circular
  non-interacting density of states with $W=1$
  (Eq.~\eqref{ASemiCirc}).
  Calculations were performed for $M=2^{11}$ bath moments.
  The phonon frequency is $\omega_0/W=0.25$,
  the coupling strength $\epsilon_p/W = 0.25$ (upper panel) and
  $\epsilon_p/W = 0.75$ (lower panel).
  Note the different scale for $\Im \Sigma(\omega)$ in the lower
  panel, where a pole of $\Sigma(\omega)$ separates the lowest band.
  The inset displays a magnification of the lowest polaron band and the pole.
}
\label{Fig:Holstein}
\end{figure}

In Fig.~\ref{Fig:Holstein} we show $A(\omega)$ for two sets of
parameters.
For $\epsilon_p/W = 0.75$, a polaron has formed as a new
quasiparticle,
which results in several separated bands in $A(\omega)$.
We also show the imaginary part of the DMFT selfenergy $\Sigma(\omega)$.
In our example, it can be obtained 
in the form of a Green function 
\begin{equation}\label{Sigma}
\Sigma(\omega) =
\langle\mathrm{vac}|cb (\omega-H_1)^{-1} b^\dagger
c^\dagger|\mathrm{vac}\rangle \;,
\end{equation}
where $H_1=(1-P_0) H (1-P_0)$ is projected onto the subspace
orthogonal to the boson vacuum $|\mathrm{vac}\rangle$, i.e.
$P_0 = |\mathrm{vac}\rangle \langle \mathrm{vac}|$.
The projection guarantees that only irreducible diagrams
contribute to $\Sigma(\omega)$.
We calculate $\Sigma(\omega)$ once $A(\omega)$ is converged.
When a polaron has formed, the polaron bands are separated by a
pole in $\Sigma(\omega)$. Similar to
Fig.~\ref{Fig:Impurity3}, a pole close to a band has to be resolved,
which requires high resolution.
Note that $\Im \Sigma(\omega)$ is zero for the lowest polaron band,
where emission of a virtual phonon is energetically forbidden.

\begin{figure}
  \includegraphics[width=0.5\textwidth]{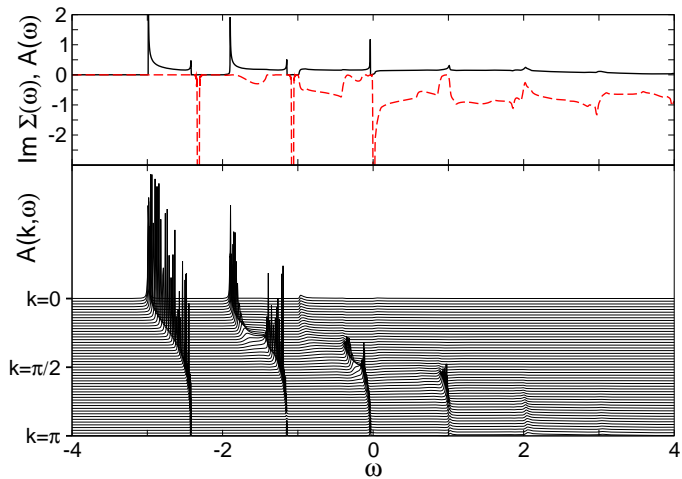}
\caption{
(Color online)
  Spectral function $A(\omega)$ for the Holstein model on a 1d chain
  ($W=4t$ with $t=1$) within DMFT-approximation,
  for $\omega_0/W=0.25$, $\epsilon_p/W=0.5$.
  The lower panel shows the $k$-resolved spectral function
  $A(k,\omega) = A^0(k,\omega-\Sigma(\omega))$,
  where $A^0(k,\omega)=\delta(\omega-\epsilon_k)$ with
  $\epsilon_k = -2t \cos k$ is the spectral function of a
  1d tight-binding chain,
  and $\Sigma(\omega)$ the DMFT selfenergy (Eq.~\ref{Sigma}).
  Calculations were performed for $M=2^{11}$ bath moments.
}
\label{Fig:Holstein2}
\end{figure}

In Fig.~\ref{Fig:Holstein2} we show a calculation for a 1d chain.
Since DMFT is constructed in the limit of high dimension,
only the basic features of polaron formation in 1d are correctly described.
With CSM, this calculation could be extended to a cluster of bosonic
sites, using one of the recently developed cluster
extensions~\cite{MJPH05} of DMFT.

\subsection{Comparison to the analytical solution}

For the Hamiltonian~\eqref{HamHolOne},
hence for the DMFT-solution of the Holstein model~\eqref{HamHol},
an explicit solution for the spectral function $A(\omega)$
can be obtained as a continued fraction~\cite{CPFF97,Su74} (CF)

\begin{equation}\label{ACF}
  A(\omega)=\frac{1}{\omega-t^2
    A_B(\omega)-\cfrac{\epsilon_p\omega_0}{\omega-\omega_0-t^2
      A_B(\omega-\omega_0) - \cfrac{2\epsilon_p\omega_0}{\dots}}} \;.
\end{equation}

We can construct the CF from a formal Lanczos recursion.
The recursion starts with the state
$c^\dagger|\mathrm{vac}\rangle$
and consecutively produces the states
$(b^\dagger/\sqrt{n!}) c^\dagger|\mathrm{vac}\rangle$,
which form an orthonormal basis of the Hilbert space $\mathcal{H}_S$
of the bosonic impurity site.
In this basis, the matrix of $H$ (Eq.~\eqref{HamHolOne}) is tridiagonal,
and $A(\omega)$, which is the $(1,1)$-element of this matrix, 
can be expressed as a CF.

To treat the coupling to the bath in Eq.~\eqref{HamHolOne},
we use Eq.~\eqref{AConvolution}.
If the electron is in the bath,
the states of the bosonic site evolve
by the Hamiltonian $H_S'=\omega_0 b^\dagger b$.
In the $n^{th}$ level of the CF~\eqref{ACF},
which corresponds to the excitation of $n$ bosons,
we must therefore insert $A_B(\omega)$ with energy shift
$A_B(\omega-n\omega_0)$.

For this example, the (Lanczos) recursion leading to the CF creates every
eigenstate of $\omega_0 b^\dagger b$ one by one. 
Otherwise, if eigenstates were mixed during the recursion, linear
combinations  $\sum_n w_n A_B(\omega-n\omega_0)$ would occur instead
of $A_B(\omega-n\omega_0)$.
The weight $w_n$ had to be determined during the recursion, and
depended on the parameters.
This is one of the many obstructions that prohibit a generalization of
Eq.~\eqref{ACF} to other models,
e.g. with a cluster of interacting (bosonic) sites or different
electron-phonon coupling.
Within CSM these consideration are pointless,
as its application does not rely on our ability to
solve one part of the Hamiltonian prior to the actual calculation.
Note also, that the general CS construction allows to calculate the
groundstate and correlation functions that cannot be expressed in a
form like~\eqref{ACF}.

\section{Time evolution of a wave packet on a chain coupled to leads}\label{Sec:Time}

In the preceding sections we used the CS construction for
the calculation of spectral properties within CSM.
It is an essential advantage of Chebyshev techniques, and the CS
construction, that they can be easily adapted to new problems.
In this section we treat the time evolution of a wave packet on a
chain coupled to leads.
To describe the coupling to a lead we can use the CS construction
from Sec.~\ref{Sec:General} without change.

Within Chebyshev techniques, time evolution of a vector 
$|\psi(t)\rangle = e^{-\mathrm{i}Ht} |\psi(0)\rangle$
is determined from the Chebyshev expansion of the time evolution
operator~\cite{TK84}
\begin{equation}\label{Time1}
  e^{-\mathrm{i}Ht} = c_0 + 2 \sum_{n=1}^\infty c_n T_n(H) \;.
\end{equation}
The expansion coefficients are given by Bessel functions 
\begin{equation}\label{Time2}
  c_n = \int_{-1}^1 \frac{T_n(x) e^{-\mathrm{i}xt}}{\pi\sqrt{1-x^2}}
  = (-\mathrm{i})^n J_n(t) \;,
\end{equation}
where $J_n(t)$ denotes the Bessel function of order $n$.
As usually, we omit the scaling of $H$.
Since $J_n(t)$ decays rapidly for $n \gg t$, 
the infinite series can be truncated to obtain $|\psi(t)\rangle$ with
high precision.
An adequate choice for the number of moments
is given by $N \gtrsim 1.5 t$.

We apply the Chebyshev time evolution to the propagation of an
electron along a chain of length $L$, which is coupled to a bath at
site $L$. The chain geometry is identical to the example studied in
Sec.~\ref{Sec:Chain}, and the Hamiltonian is given by
Eq.~\eqref{HamChain}.
The calculation of $T_n(H)|\psi(0)\rangle$ proceeds along the lines
established in previous sections,
and $|\psi(t)\rangle$ is obtained from
Eqs.~\eqref{Time1},~\eqref{Time2}.

\begin{figure}
\includegraphics[width=0.5\textwidth]{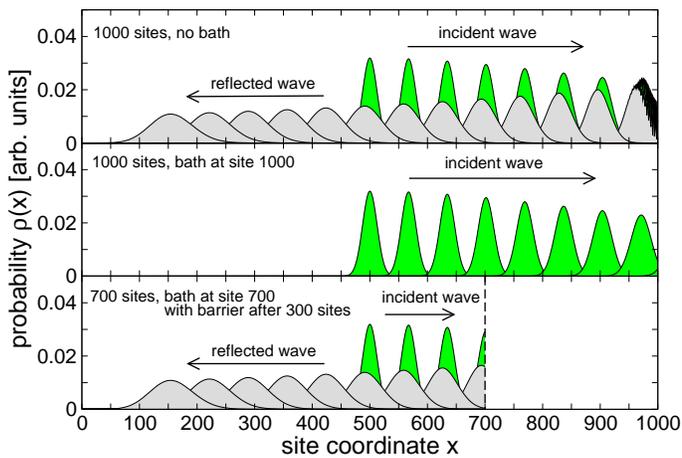}
\caption{
(Color online)
Time evolution of a Gaussian wave packet
  Eq.~\eqref{WavePacket}, with $2\sigma^2 = 25^2$ and $K=1$.
  The time step between two curves is $\delta t=40$.
  The group velocity of the wave packet is given by the dispersion
  $\epsilon_k = -2 t_\mathrm{hop} \cos k$ as
  $v = (\partial \epsilon_k / \partial k)|_K$.
  Per time step, the wave packet moves $v \times \delta t \approx 67$ lattice
  sites.
  The three panels correspond to the situations described in the text.}
\label{Fig:Time}
\end{figure}

At $t=0$, we place a Gaussian wave packet of width $\sigma$ and
momentum $K$ centered at site $m_0=L/2$,
\begin{equation}\label{WavePacket}
  |\psi(0)\rangle \propto \sum_m e^{\mathrm{i}Km} e^{-(m-m_0)^2/2\sigma^2}
  c^\dagger_m |\mathrm{vac}\rangle \;,
\end{equation}
(we omit normalization here),
and let it evolve in time (see Fig.~\ref{Fig:Time}).
Without a bath,
the particle is reflected at the right end of the chain and returns,
moving to the left.

We now add a bath to the right end,
whose spectral function $A_B(\omega)$ is given by
Eq.~\eqref{ASemiCirc}, with $W=4t$.
Since this is the spectral function of a half-infinite
chain at its open end, also the full Hamiltonian $H$ describes a
half-infinite chain.
Physically, the open boundary at site $L$ is removed by coupling to an
infinite lead.
In contrast to the previous situation, the particle will not be
reflected at site $L$, but propagates into the bath (or lead).
As the middle panel in Fig.~\ref{Fig:Time} shows, no
spurious reflection occurs.

In the Hamiltonian $H$, sites $1 \dots L$ are explicitly included
while sites $\ge L+1$ of the lead are realized through the bath.
In this way, we can use our Chebyshev approach to realize
`transparent boundary conditions'~\cite{HF94,Mo03} that mimic an infinite
system with a finite number of lattice sites explicitly treated. 
Transparent boundary conditions
are implemented in Refs.~\onlinecite{HF94,Mo03} by a modification
of time propagation algorithms like e.g. Crank-Nicholson.
Within the CS construction, modified boundary conditions
are implemented independently of the actual calculation.
For time propagation we use the same
Eqs.~\eqref{ChebyImpurity1}--\eqref{ChebyImpurity3} as for the
calculation of spectral properties.
This permits us to use a variety of algorithms,
like we used Lanczos algorithm in Sec.~\ref{Sec:Bosonic}.
We could also use the CS in combination with
Crank-Nicholson, but for the time-independent Hamiltonians considered here
Chebyshev time propagation is most efficient.

Transparent boundary conditions correspond to a particular choice for
$A_B(\omega)$.
We have the freedom to choose any $A_B(\omega)$.
For example, if $A_B(\omega)$ is the spectral function of a $300$ site
chain at its end, and $L=700$, $H$ describes a chain of length $L+300=1000$.
As before, the particle is reflected and returns moving to the left
(see bottom panel in Fig.~\ref{Fig:Time} in comparison to the top panel).
The point is that the reflecting barrier is now realized through a
bath with carefully chosen $A_B(\omega)$, and not by 
lattice sites $701\dots L$ explicitly included in the Hamiltonian.
We might call these `reflecting boundary conditions'.

\begin{figure}
\includegraphics[width=0.5\textwidth]{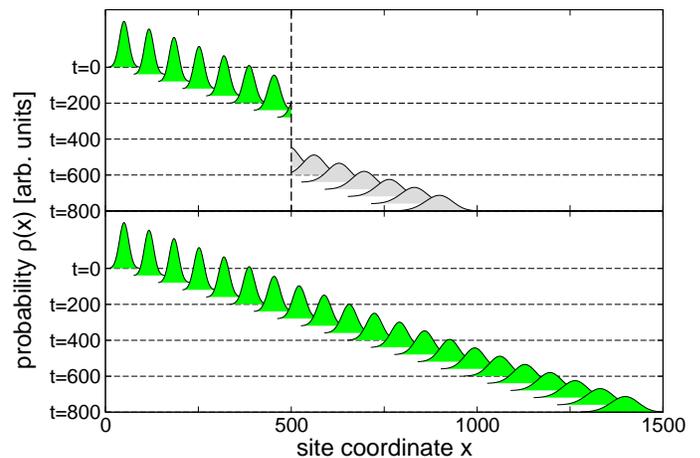}
\caption{
(Color online)
Time evolution of a Gaussian wave packet
with same parameters as in Fig.~\ref{Fig:Time}.
Time is shown on the vertical axis.
A bath which mimics a chain segment of $500$ sites is placed between
sites $500$ and $501$ of the chain (vertical dashed line).
The wave enter the bath and reemerges after a delay $\Delta t = 500/v
\approx 300$ (upper panel).
After transmission through the bath the wave packet is $500$ sites
behind the wave packet on a simple chain,
whose propagation is shown in the lower panel for comparison.}
\label{Fig:Time2}
\end{figure}

For Fig.~\ref{Fig:Time2}
we slice the chain in two parts which are connected by a bath.
Similar to the previous example, the bath shall replace a finite
segment of the chain.
Since the bath has now two entry points  -- at its junction to the
left and to the right part of the chain --,
this example involves off-diagonal bath spectral functions.
Let $d^{(\dagger)}_{1/2}$ denote the operators to the two entry
points.
Following Sec.~\ref{Sec:General},
we need two different types of Chebyshev vectors
$|n_b\rangle = T_n(H_B) d^\dagger_b |\mathrm{vac}\rangle$ for $b=1,2$.
Eq.~\eqref{ChebyImpurity3} generalizes to
$ d_a |n_b\rangle = \mu^{B,ab}_n |\mathrm{vac}\rangle$,
where the four different types of moments $\mu^{B,ab}_n$
correspond to the four different spectral function characterizing the
bath: The diagonal $A^B_{11}(\omega)$, $A^B_{22}(\omega)$ for moving into
the bath and back, and the off-diagonal $A^B_{12}(\omega)$, $A^B_{21}(\omega)$
for moving through the bath.
The latter ones give the energy-dependent transmission rate through
the bath.
Only for non-zero $A^B_{12}(\omega)$, $A^B_{21}(\omega)$,
the bath connects the two parts of the chain.
In Fig.~\ref{Fig:Time2} the spectral functions of the bath are these
of a finite chain segment. We see how the wave moves
into the bath at its left side, passes through and reemerges at the
right site. 
The bath perfectly mimics the transmission through a chain segment.
As we noted above, we have the freedom
to change the behavior by a different choice of the bath spectral
functions $A^B_{ab}(\omega)$.

\section{Summary and outlook}\label{Sec:Summary}

In this article we introduced the Chebyshev space method (CSM)
for the treatment of degrees of freedom with non-trivial dynamics.
In our examples the degree of freedom is
given by the operator $d^\dagger$ which creates a fermion in a bath,
and the dynamics of $d^\dagger$ is specified by a spectral function
$A_B(\omega)$.
We demonstrated for various examples, how CSM yields extremely accurate
results with modest computational effort.
For the example in Secs.~\ref{Sec:NumRes},~\ref{Sec:Bosonic}
the computational effort is even independent of the geometry or dimension.

A particular advantage of the Chebyshev space (CS) construction is that it 
still provides a Hamiltonian $H_B$ acting on a Hilbert space,
albeit it is an abstract space without a direct physical counterpart.
We can use this form of $H_B$ in different situations,
like in the Lanczos algorithm or for time propagation.
We have therefore obtained a numerically exact Hamiltonian
treatment of degrees of freedom with non-trivial dynamics,
which does not rely on a discretization of spectral functions.
To the best of our knowledge, this has been achieved for the first
time.

The concept of a CS and an ancillary Hamiltonian
$H_B$ acting on this space is very general.
It makes no assumption about the Hamiltonian $H_S$ for
the quantum system, and can be extended to various problems mentioned in
the introduction.
We have obtained preliminary results for the Holstein model within
cluster extensions of DMFT,
but the issue needs further exploration from the physical point
of view.

An especially important topic is the extension to
interacting fermions at finite density, or to bosonic baths.
With a single fermion, the bath is in the vacuum state
$|\mathrm{vac}\rangle$ if the fermion is removed by application of the
operator $d$.
This simplification is no longer true for finite fermion density,
when adding and removing a fermion creates particle-hole
pairs in the bath, which initially was prepared as the particle-hole
vacuum (or Fermi sea).
We have derived a formalism that deals with particle-hole
pairs in the context of CSM.
The CS construction is used in this formalism without change. 
A similar formalism for bosonic baths can be derived.
As a first result, this formalism allows for the calculation of
the core level spectral function in the X-ray absorption problem~\cite{Mah00}.
The spectral function can be calculated exactly in this case.
It is equivalent to the Anderson model with one immobile spin species,
but the extension to the full Kondo problem is much harder. 
The presentation of this formalism is left for a future publication.

The CS construction can be also used in the context of
(diagrammatic) Green function techniques.
Any Feynman diagram contains energy-dependent Green functions which
represent a degree of freedom with non-trivial dynamics.
The difficult task is to sum up a huge number of Feynman diagrams.
Note that we can interpret the self-consistent calculation of the
Holstein model in Sec.~\ref{Sec:Holstein} as the exact
summation of a certain class of Feynman diagrams. 
The bosonic impurity model defines a set of skeleton diagrams for this
problem, and imposing self-consistency corresponds to replacing bare
Green functions (`thin lines' in a diagram) by renormalized Green
functions (`thick lines').
In this example, diagrams are selected by a geometric rule,
but with some modifications a different selection rule can be implemented.
The CSM provides the link between exact numerical techniques for
Hamilton operators, and general approximation schemes
for Green functions.

In conclusion, we believe that the CSM introduced in this work 
is a powerful addition to existing numerical techniques in
theoretical physics or chemistry.
It substantially enlarges the field of applications of Chebyshev
techniques and keeps their advantages.
The results we obtained in this work are promising for successful
applications to more complicated problems,
and the possible combination of CSM with other techniques shall
prove fruitful.
The further development of CSM and its application to the study of
physical problems mentioned in the introduction is the subject of
current research.

\textit{Acknowledgment.} We are indebted to G. Wellein for helpful
comments during the preparation of this manuscript.

\appendix

\section{Chebyshev expansions and the kernel polynomial}\label{App:Cheby}

In this article we repeatedly approximate a function $f : [-1,1] \to
\mathbb{R}$ by a finite Chebyshev series
\begin{equation}\label{fChebyFinite}
  f_N(x) = \frac{1}{\pi \sqrt{1-x^2}} \Big[ \mu_0 + 2 \sum_{n=1}^{N-1}
  \mu_n T_n(x) \Big] \;,
\end{equation}
where the moments $\mu_n$ are given by
\begin{equation}\label{fChebyMoms}
  \mu_n = \int_{-1}^1 f(x)T_n(x) \, dx \;.
\end{equation}
A central question is how good $f(x)$ is approximated by $f_N$.
Functional analysis teaches us that the $f_N$ converge to $f$ in the
integral norm 
\begin{equation}
\|f(x)-f_N(x)\| = \left[\int_{-1}^1 \frac{(f(x)-f_N(x))^2} {\pi \sqrt{1-x^2}} \, dx
\right]^{1/2}
\end{equation}
corresponding to the scalar product used in~\eqref{ChebyOrthogonal},
provided $\|f(x)\|$ is finite.
For practical purposes, this convergence property is too weak,
and we demand uniform or at least pointwise convergence.
We do not try to resolve the issue of convergence here,
but state that for sufficiently smooth $f(x)$ 
the series $f_N$ will converge uniformly to $f$ on any closed
sub-interval of $[-1,1]$ that excludes the end-points~$\pm 1$.
Convergence at the end-points is of course spoiled by the divergence of the
weighting function $(1-x^2)^{-1/2}$.

In physical applications, we want to expand spectral functions which
are not necessarily continuous, and perhaps contain some $\delta$-peaks due to
(quasi-) particle states with infinite lifetime.
In this case, the series cannot converge uniformly  -- otherwise its
limit $f$ had to be continuous itself --, but the situation is even worse.
The infamous Gibbs phenomenon ruins convergence at all:
In plain words, the series~\eqref{fChebyFinite} fails to converge in
the vicinity of a discontinuity, e.g. a jump of the function,
instead shows rapid oscillations whose height does not decrease
as the number of terms $N\to\infty$.
The Gibbs phenomenon is a severe obstacle for practical applications.

\begin{figure}
\includegraphics[width=0.5\textwidth]{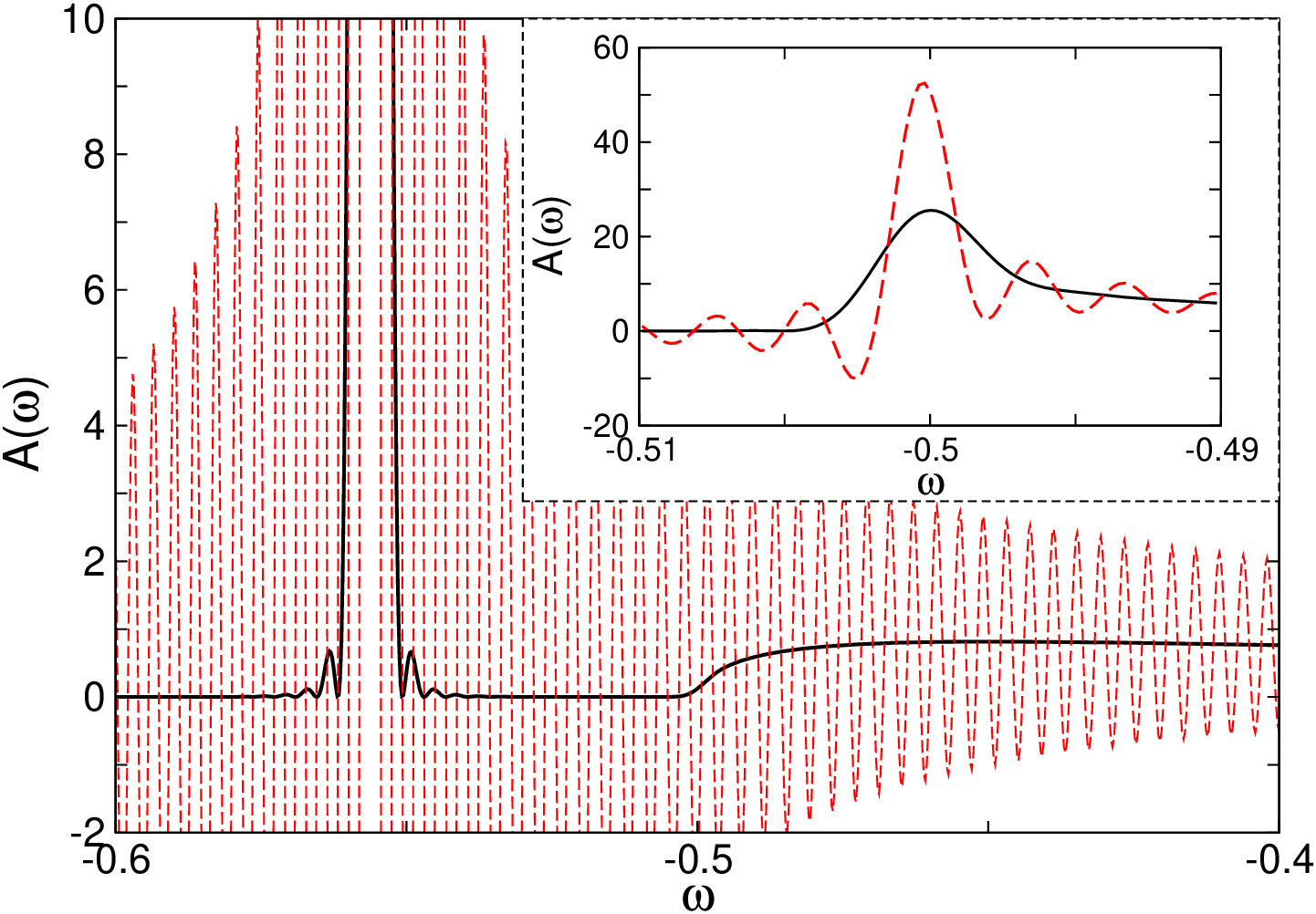}
\caption{(Color online) Shown is the spectral function to the
  Hamiltonian~\eqref{HOneSite} and $\Delta=0.4$.
  The solid curve is
  calculated with attenuation factors from Eq.~\eqref{JacksonFactors}
  and is identical to the curve in Fig.~\ref{Fig:Impurity3}.
  The dashed curve is calculated without attenuation factors.
  The inset displays the curve from Fig.~\ref{Fig:Impurity3} to
  $\Delta=0.26$ and $N=2^{10}$,
  again with (solid) and without (dashed) 
  attenuation factors.
  These two curves illustrate the remark from the text:
  After multiplication with attenuation factors
  the reconstructed spectral function seems to be smooth.
  But the unmodified moments correspond to a discontinuous spectral
  function with a pole close to $\omega=-0.5$.
  The slow decay of moments of this spectral function results in oscillations.
  The pole is resolved only by increasing $N$ and keeping
  the attenuation factors (cf. the curves in Fig.~\ref{Fig:Impurity3} up
  to $N=2^{16}$). }
\label{Fig:Gibbs}
\end{figure}

Fortunately, the problem arising from the Gibbs phenomenon is solved for
Chebyshev expansions~\cite{SR94,WZ94,Wa94,SRVK96}.
It is possible to specify, prior to the calculation of the $\mu_n$,
a set of attenuation factors $g_n^N$, $n=0,\dots,N-1$ for every $N$,
such that the modified approximants 
\begin{equation}
  \tilde{f}_N = \frac{1}{\pi \sqrt{1-x^2}} \Big[g^N_0 \mu_0 + 2 \sum_{n=1}^{N-1}
  g^N_n \mu_n T_n(x) \Big]
\end{equation}
do not show the Gibbs phenomenon but approximate a wide class of
functions in a favourable way.
Plainly spoken, the attenuation factors damp out high frequency
oscillations that otherwise lead to spurious results in the
finite series. 
Instead of a more precise formulation we show an example in Fig.~\ref{Fig:Gibbs}.
The best choice in many cases are factors $g_n^N$ derived from the so-called
Jackson kernel~\cite{SRVK96}, leading to the explicit expression
\begin{equation}\label{JacksonFactors}
  g_n^N =  \frac{(N-n+1)\cos\frac{\pi n}{N+1} 
    + \sin\frac{\pi n}{N+1}\cot\frac{\pi}{N+1}}{N+1} \;.
\end{equation}
Modifying moments with these attenuation factors is related to
convolution with an almost Gaussian peak of width
$\sigma = \pi/N$ (cf. Ref.~\onlinecite{WWAF06}).
An important property of the modification with these attenuation
factors is that the modified approximants $\tilde{f}_N$ are positive
whenever the function $f$ is.
This is not true for the unmodified approximants
(cf. Fig.~\ref{Fig:Gibbs}).
Besides Eq.~\eqref{JacksonFactors}, other suggestions for attenuation factors exist,
see Ref.~\onlinecite{WWAF06} for a different choice in the context of Green
functions, with an application in Ref.~\onlinecite{hwbaf06}.

Note that the factors $g_n^N$ in Eq.~\eqref{JacksonFactors} have the
property that $g_n^N \to 1$ for $N\to\infty$.
We can therefore always use the attenuation factors $g_n^N$
without losing information.
For functions with discontinuities 
multiplication with attenuation factors enforces
the decay of moments relevant for good approximation properties.
For smooth $f(x)$ the moments itself decay rapidly,
and multiplication with the attenuation factors does not change the
result for not too small $N$.

The introduction of
attenuation factors is the essence of the kernel polynomial method (KPM).
Within KPM resolution for spectral functions 
is already obtained with a fairly small number of moments,
thus allowing for accurate calculations with moderate demands on
computational time or memory.
The possible resolution is much better than known e.g. from the
Lanczos algorithm where peaks are commonly broadened by a Lorentzian
function. 
Furthermore, the resolution of the Chebyshev expansion is uniform over
the energy interval, and does not deteriorate towards the center of
the spectrum.

\section{Determinants and two-term recurrences}\label{App:Recurrence}

We formulate a well-known determinant identity related to two-term
recurrences. Define, for each $n$ and given $a_j, b_j, c_j$,  a tridiagonal $n\times n$-matrix
\begin{equation}
  A_n = \begin{pmatrix}
    a_1 & c_1 &  \\
    b_1 & a_2 & c_2 \\
       & b_2 & a_3  & \ddots \\
& & \ddots & \ddots & \ddots \\    
& & & \ddots &    a_{n-2} & c_{n-2} &  \\
& & & &    b_{n-2} & a_{n-1} & c_{n-1} \\
& & & &     & b_{n-1} & a_n \\
\end{pmatrix} \;.
\end{equation}
Expanding in the last row yields
\begin{equation} 
  \det A_n = a_n \det A_{n-1} - b_{n-1} c_{n-1} \det A_{n-2} \;.
\end{equation}

From this relation it follows that the characteristic polynomials
$\chi_n = \det(x-A_n)$ obey the two-term recurrence 
\begin{equation}\label{App:CharPoly}
\begin{split}
  \chi_0 &= 1 \,,\; \chi_1 = x-a_1 \,,\; \\
  \chi_{n+1} &= (x-a_{n+1}) \chi_{n} - b_n c_n \chi_{n-1} \;.
\end{split}
\end{equation}

Conversely, for polynomials $P_n$ defined by a two-term recurrence 
\begin{equation}
 P_0 = 1 \,,\; P_1 = a_1 x \;, P_n = a_n x P_{n-1} - b_{n-1} P_{n-2} \;,
\end{equation}
this result implies that
\begin{equation}
  P_n = \det  \begin{pmatrix}
a_1 x & d_1 \\ c_1 & a_2 x & d_2 & & \\ & c_2 & a_3 x
    & d_3 \\ & & c_3 & a_4 x \\ & & & & \ddots    
\end{pmatrix} \;,
\end{equation}
where $c_n$, $d_n$ is chosen in such a way that $c_n d_n = b_n$,
e.g. for positive $b_n$ we may choose
$c_n=d_n=\sqrt{b_n}$.
By multiplying the $j^{th}$ column -- or equivalently, $j^{th}$ row -- by
$1/a_j$, $P_n$ can be expressed as the characteristic polynomial of a
tridiagonal matrix times a prefactor $\prod_{j=1}^n a_j$.

In particular, the Chebyshev polynomials fulfill
\begin{equation}
  T_n(x) = \det \begin{pmatrix}
x & 1 \\ 1 & 2x & 1 & & \\ & 1 & 2x
    & 1 \\ & & 1 & 2x \\ & & & & \ddots    
  \end{pmatrix} \;.
\end{equation}
Multiplying the second to the last column by $1/2$,
we find the result $T_M(x)=2^{M-1} \det (x-H_B^M)$
used in the main text.

\section{Hermiticity of $H_B$}\label{App:Herm}

We first prove that $H_B$ is hermitian by calculating explicitly
that $H_B$ is equal to its adjoint.
The product of two Chebyshev polynomials
satisfies the identity~\cite{AS70}
\begin{equation}\label{ChebyProduct}
  T_m(x) T_n(x) = (T_{m+n}(x)+T_{m-n}(x))/2 \;.
\end{equation}
For notational convenience, we define here $T_{-n}(x) = T_n(x)$.
It follows for Chebyshev vectors $|n\rangle=T_n(H_B)|0\rangle$
that
\begin{equation}
  \langle m|n \rangle= (\mu^B_{m+n} + \mu^B_{m-n})/2 \;,
\end{equation}
where we again define $\mu^B_{-n} = \mu^B_n$.
The scalar product of two vectors $|a\rangle = \sum_{n=0}^\infty a_n
|n\rangle$, $|b\rangle = \sum_{n=0}^\infty b_n |n\rangle$ given as linear
combinations of Chebyshev vectors is now found as 
\begin{equation}\label{ChebyScal}
  \langle b|a \rangle = \sum_{m,n=0}^\infty b_m^* a_n (\mu^B_{m+n} +
  \mu^B_{m-n})/2 \;.
\end{equation}
Since $\mu^B_{m-n}=\mu^B_{n-m}$, with the notational convention adopted
above, we have $\langle a|b \rangle =  \langle b|a \rangle^*$, as it
must be (note that the moments are real).

To calculate a scalar product with $H_B$
we use that
\begin{equation}
  H_B|a\rangle = \frac{1}{2} \sum_{n=0}^\infty a_n (|n-1\rangle+|n+1\rangle) \;,
\end{equation}
where the summand for $n=0$ is 
$a_0(|-1\rangle+|1\rangle)/2 = a_0 |1\rangle$ 
in accordance with Eq.~\eqref{HBTruncated}.
Inserting this equation we find
\begin{equation}\label{HScalar}
\begin{split}
  \langle b|H_B|a\rangle &= \frac{1}{2} \sum_{m,n=0}^\infty b_m^* a_n (\langle
  m|n-1\rangle + \langle m|n+1 \rangle)  \\
  &=  \frac{1}{4} \sum_{m,n=0}^\infty b_m^* a_n
  (\mu^B_{m+n-1}+\mu^B_{m-n+1} \\ &\qquad\quad
  +\mu^B_{m+n+1}+\mu^B_{m-n-1} ) \;.
\end{split}
\end{equation}
We can condense this equation to matrix-vector form
\begin{equation}
  \begin{split}
    \langle b|H_B|a\rangle &= \frac{1}{4} \sum_{m,n=0}^\infty b_m^* a_n M_{mn} \;, \\
    M_{mn} &= \mu^B_{m+n-1}+\mu^B_{m-n+1}+\mu^B_{m+n+1}+\mu^B_{m-n-1} \;.
\end{split}
\end{equation}
Since $M_{mn}$ depends only on the sum and difference of $m,n$,
$M_{mn}=M_{nm}=M_{nm}^*$ (the moments are real, and $\mu^B_{-l}=\mu^B_l$).
It follows that $\langle b|H_B|a\rangle= \langle a|H_B|b\rangle^*$, and
$H_B$ is hermitian.

We derived the hermiticity of $H_B$ without imposing any constraint on the
moments.
This may be a bit puzzling, because the moments should correspond to a
non-negative spectral function $A_B(\omega)$ if $H_B$ is hermitian.
The puzzle is resolved by observing that the scalar
product~\eqref{ChebyScal} has to be positive (semi-) definite.
This requirement imposes constraints on the moments that are equivalent to a
non-negative $A_B(\omega)$.
To mention one, $\langle n|n \rangle = (\mu^B_0 + \mu^B_{2n})/2$, so it must
$\mu^B_{2n}\ge -\mu^B_0$.
Consider for example $H_B=0$, equivalent to
$A_B(\omega)=\delta(\omega)$. We then have $\mu^B_{2n}=(-1)^n$,
$\mu^B_{2n+1}=0$, i.e. equality holds in this inequality. The scalar
product~\eqref{ChebyScal} is positive semi-definite, but not positive
definite, as $|n\rangle=0$ and $\langle n|n\rangle=0$ for odd $n$.

We can characterize the properties of the scalar product~\eqref{ChebyScal}
better, 
if we consider Chebyshev expansions 
\begin{equation}
f(x) = \sum_{m=0}^\infty f_m T_m(x) \;, \quad g(x) = \sum_{m=0}^\infty g_m
T_m(x) 
\end{equation}
of functions $f,g: [-1,1] \to \mathbb{R}$,
and the associated vectors 
\begin{equation}
|f\rangle = \sum_{m=0}^\infty f_m |m\rangle \;, \quad
|g\rangle = \sum_{m=0}^\infty g_m |m\rangle
\end{equation}
in $\mathcal{H}_c$.
With Eqs.~\eqref{ChebyProduct},~\eqref{ChebyMoms}
we find for $A_B(\omega)$ as in Eq.~\eqref{ACheby}
$\int_{-1}^1 T_m(x) T_n(x) A_B(x) dx=(\mu^B_{m+n}+\mu^B_{m-n})/2$,
which implies
\begin{equation}\label{ScalarAB}
  \int_{-1}^1 f(x) g(x) A_B(x) \, dx = \langle f|g\rangle \;.
\end{equation}
The scalar product~\eqref{ChebyScal} in $\mathcal{H}_c$ 
therefore corresponds to a
scalar product of functions $[-1,1]\to\mathbb{R}$, given as the
integral on the left hand side of this equation.
We conclude, that for continuous $A_B(\omega)$ the scalar
product~\eqref{ChebyScal} is positive semi-definite (positive
definite) if and only if $A_B(\omega) \geq 0$ (and $A_B(\omega)$ does
not vanish on an open interval).

We discussed in App.~\ref{App:Cheby} that a truncated, finite Chebyshev
expansion of $A_B(\omega)$ may fail to be positive although
$A_B(\omega)$ is.
Positivity of the finite Chebyshev expansion is ensured by
using attenuation factors of a positive kernel like in
Eq.~\eqref{JacksonFactors}.
To obtain a positive definite scalar product~\eqref{ChebyScal}
for a finite Chebyshev expansion we must therefore use modified
moments $\mu^B_n g^M_n$ instead of unmodified moments $\mu^B_n$.
For the Lanczos algorithm, used in Sec.~\ref{Sec:Bosonic},
which aims at constructing an orthonormal basis in Krylov-subspaces,
positive definiteness of~\eqref{ChebyScal} is crucial.
For the calculation of spectral or dynamical properties 
it is not that essential.
But we learned in App.~\ref{App:Cheby} that the modification of
moments by attenuation factors never ruins a result.
Hence our recommendation is to always modify 
the moments $\mu^B_n$ put into the calculation
with attenuation factors.

\section{Eigenstates of $H_B^M$}\label{App:Eigen}

According to App.~\ref{App:Recurrence}
the characteristic polynomial of $H_B^M$ (Eq.~\eqref{HBTruncated})
is $\det[x-(H^M_B)_{mn}] = 2^{-(M-1)} T_M(x)$.
The $M^{th}$ Chebyshev polynomial $T_M(x)=\cos(M \arccos x)$
has $M$ distinct real roots 
\begin{equation}
  x_j= \cos \frac{\pi (j-1/2)}{M} \;, \quad   j=1,\dots,M \;.
\end{equation}
We concluded in Sec.~\ref{Sec:TruncatedHB} that $H_B^M$ is
diagonalizable with real eigenvalues $x_j$.
The eigenstates of $H_B^M$ can be given explicitly, using the
recurrence~\eqref{ChebyRecursion}.
The state
\begin{equation}\label{HBStates}
  |\phi_j\rangle =  |0\rangle + 2 \sum_{m=1}^{M-1} T_m(x_j) |m\rangle
\end{equation}
is the eigenstate of $H_B^M$ to eigenvalue $x_j$, i.e.
$H^M_B |\phi_j\rangle = x_j |\phi_j\rangle$.
If we use Eq.~\eqref{HBTruncated} to calculate $H^M_B |\phi_j\rangle$,
we find that the term proportional to $T_M(x_j)$ drops out, because $x_j$ is a
root of $T_M(x)$. This explains why the roots of $T_M(x)$ occur as
eigenvalues of $H^M_B$.

Scalar products of the states $|\phi_j\rangle$ have to be calculated using
Eq.~\eqref{ChebyScal}. 
It turns out, that the $|\phi_j\rangle$ are not orthogonal to each
other. Indeed, while $H_B$ is hermitian, the truncated $H^M_B$ is not
hermitian. 
If one repeats the calculations leading to Eq.~\eqref{HScalar},
now with upper summation boundary $M-1$ instead of $\infty$,
one finds that a term $a_{M-1} |M\rangle$ is missing, and the scalar
product is not symmetric in $|a\rangle$, $|b\rangle$.
The mathematical reason that $H^M_B$ fails to be hermitian, is that
$H^M_B$ is obtained from $H_B$ via a non-orthogonal projection.
Working with a non-hermitian operator could in principle spoil the
calculation for $M<N$.
We found in Sec.~\ref{Sec:TruncatedHB} that this does not happen.
By an explicit calculation of the spectral function encoded by
$H_B^M$, we now show that the non-hermiticity of $H_B^M$ has no
(negative) consequences.

As Fig.~\ref{Fig:Scaling} shows, for $N \gg M$ 
CSM resolves $M$ poles in the spectral function encoded by $H^M_B$.
The position of the poles is determined by $x_j$,
and their weight $w_j$ can be calculated using the eigenstates
$|\phi_j\rangle$.
We first expand the $0^{th}$ Chebyshev vector $|0\rangle$ in the
$|\phi_j\rangle$.
With the identity
\begin{equation}\label{ChebyDiscrete}
 \frac{1}{M} \sum_{j=1}^M T_m(x_j) T_n(x_j) = \begin{cases}
    \frac{1}{2} \, \delta_{mn} \;, \quad &  m,n \ne 0 \\
    1 \;, &  m=n=0
  \end{cases} \;,
\end{equation}
which is a kind of discrete orthogonality relation for Chebyshev
polynomials, we find that
\begin{equation}
  |0\rangle = \frac{1}{M} \sum_{j=1}^M |\phi_j\rangle \;.
\end{equation}
The coefficients in this linear combination are $T_0(x_j)=1$.
Now $\delta(\omega-H^M_B) |\phi_j\rangle =  \delta(\omega-x_j)
|\phi_j\rangle$, and we obtain 
\begin{equation}
  \langle 0| \delta(\omega-H^M_B) |0\rangle 
  = \sum_{j=1}^M w_j \delta(\omega-x_j) \;,
\end{equation}
with the weight of each pole given by
\begin{equation}
\begin{split}\label{HBWeight}
  w_j =  \frac{1}{M}\langle 0|\phi_j\rangle &= \frac{1}{M} \Big[ \mu^B_0 + 2 \sum_{m=1}^{M-1}
  T_m(x_j) \mu^B_m \Big] \\
  &=\frac{\pi}{M} \sqrt{1-x_j^2} \, A_B(x_j) \;.
\end{split}
\end{equation}
The first line in this equation follows from the
definition~\eqref{HBStates},
the second line from comparison with Eq.~\eqref{ACheby}.
Here $A_B(\omega)$ is assumed as an expansion with $M$ Chebyshev
moments, in accordance with the truncation of $H_B^M$.
The weight $w_j$ of the pole at $x_j$ is therefore the value of
$A_B(\omega)$ at the position of the pole, weighted by the inverse
$(1-x^2)^{1/2}$ of the weighting function for Chebyshev expansions
(and with $1/M$ for $M$ poles).
Using the attenuation factors from Eq.~\eqref{JacksonFactors},
$A_B(\omega)$ is positive, which implies that the weight $w_j$ of each
pole is positive.
The total weight is
\begin{equation}
  \sum_{j=1}^M w_j = \langle 0|  \frac{1}{M} \sum_{j=1}^M
  |\phi_j\rangle = \langle 0|0\rangle = \mu^B_0 = 1 \;,
\end{equation}
according to the sum rule $\mu^B_0 = \int_{-1}^1 A_B(x) dx =  1$.
This result also follows with the explicit expression for $w_j$ 
in terms of the $\mu^B_n$ (first line of Eq.~\eqref{HBWeight}), and
Eq.~\eqref{ChebyDiscrete}.
If we express $w_j$ by $A_B(\omega)$ (second line of
Eq.~\eqref{HBWeight}),
we find a nice relation to quadrature formulas
from numerical integration.
In the expression for the total weight
\begin{equation}
  \sum_{j=1}^M w_j = \frac{\pi}{M} \sum_{j=1}^M \sqrt{1-x_j^2} \,
  A_B(x_j) \;,
\end{equation}
the right hand side is the formula for Gauss-Chebyshev
integration~\cite{PTVF92} of a function, with abscissas at $x_j$.
Now $A_B(\omega)$, expanded with $M$ Chebyshev moments,
is a polynomial of order $M$ times the weighting function
$(1-x^2)^{-1/2}$.
The Gauss-Chebyshev integration formula is exact in this case,
and we find once again $\sum_{j=1}^M w_j = \int_{-1}^1 A_B(x) dx =
1$.

We can finally understand why $H_B^M$ is non-hermitian using the scalar
product Eq.~\eqref{ScalarAB}.
A state $|\phi_j\rangle$ corresponds to the function
\begin{equation}
  \phi_j(x) = T_0(x_j) T_0(x) + 2 \sum_{m=1}^{M-1} T_m(x_j) T_m(x)
\end{equation}
which is the $M^{th}$ Chebyshev approximation to $\delta(x-x_j)$.
With a finite number of moments, $\phi_j(x)$ has finite width.
Using Eq.~\eqref{ScalarAB} for the scalar product,
\begin{equation}
  \langle \phi_j | \phi_k \rangle = \int_{-1}^1 \phi_j(x) \phi_k(x)
  A_B(x) dx \;,
\end{equation}
we see that the integral is non-zero even for $j \ne k$, since the two
functions $\phi_j(x)$, $\phi_k(x)$ have finite overlap.
The deviation of $H_B^M$ from hermiticity is therefore an effect of
finite resolution of a finite Chebyshev expansion.
The larger $M$, the smaller is the overlap and hence the scalar
product.
We conclude that in the limit $M\to\infty$, the states
$|\phi_j\rangle$ are mutually orthogonal, in accordance with
hermiticity of $H_B$ proved in App.~\ref{App:Herm}.

We have obtained two important results in this Appendix.
First,
despite its non-hermiticity, $H^M_B$ encodes a positive, normalized
spectral function.
Using the truncated $H^M_B$,
as in Sec.~\ref{Sec:TruncatedHB} for $M<N$, does therefore not lead to
erroneous results.
Second, we related the CS construction for $H_B$ to an explicit
representation Eq.~\eqref{HamBath}.
In particular, Eq.~\eqref{HBWeight} shows how to obtain
from given moments $\mu^B_m$ a discretization
(Eq.~\eqref{Aapproximation}) of $A_B(\omega)$
(the opposite way is given by Eq.~\eqref{ChebyMoms}).
The point is that the CS construction indicates how to
discretize a given $A_B(\omega)$ to prescribed resolution.
Within CSM we can work without a discretization anyway.
The reader should note that the CS construction is no longer
equivalent to the discretization of $A_B(\omega)$ in the extensions of
CSM mentioned in the Summary (Sec.~\ref{Sec:Summary}).


\end{document}